\documentclass[12pt]{article}
\usepackage{amsmath}
\usepackage{amssymb}
\usepackage{cite}
\usepackage{float}
\usepackage{latexsym}
\usepackage{color}
\usepackage{graphicx}
\usepackage{caption}
\usepackage{hyperref}
\usepackage{mathtools}
\numberwithin{equation}{section}
\newcommand{\beq}{\begin{equation}}
\newcommand{\be}{\begin{equation}}
\newcommand{\ee}{\end{equation}}
\newcommand{\bea}{\begin{eqnarray}}
\newcommand{\eea}{\end{eqnarray}}

\newcommand{\pa}{\partial}

\newcommand{\nn}{\nonumber}

\newcommand{\mc}{\mathcal}

\begin{document}

\pagenumbering{Alph}
\begin{titlepage}
\hbox to \hsize{\hspace*{0 cm}\hbox{\tt }\hss
    \hbox{\small{\tt }}}

\vspace{1 cm}

\centerline{\bf \large Larger Twists and Higher $n$-Point Functions }

\vspace{.6cm}

\centerline{\bf \large  with Fractional Conformal Descendants}

\vspace{.6cm}

\centerline{\bf \large in $S_N$ Orbifold CFTs at Large $N$}

%\vspace{.6cm} \centerline{\bf \Large }
\vspace{1 cm}
 \centerline{\large
Benjamin A. Burrington $^{\star}$\footnote{benjamin.a.burrington@hofstra.edu}\,,
A.W. Peet$^{\dagger\S}$\footnote{awpeet@physics.utoronto.ca}}

\vspace{0.5cm}

\centerline{\it ${}^\star\!\!$ Department of Physics and Astronomy, Hofstra University, Hempstead, NY 11549, USA}
\centerline{\it ${}^\dagger$Department of Physics, University of Toronto, Toronto, ON M5S 1A7, Canada}
\centerline{\it ${}^\S$Department of Mathematics, University of Toronto, Toronto, ON M5S 2E4, Canada}

\vspace{0.3 cm}

\begin{abstract}
We consider correlation functions in symmetric product ($S_N$) orbifold CFTs at large $N$ with arbitrary seed CFT, expanding on our earlier work \cite{Burrington:2022dii}.  Using covering space techniques, we calculate descent relations using fractional Virasoro generators in correlators, writing correlators of descendants in terms of correlators of ancestors.  We first consider the case three-point functions of the form ($m$-cycle)-($n$-cycle)-($q$-cycle) which lift to arbitrary primaries on the cover, and descendants thereof.  In these examples we show that the final descent relations do not depend on the covering space data, nor on the specific details of the seed CFT.  This makes these descent relations universal in all $S_N$ orbifold CFTs.  Next, we explore four-point functions of the form (2-cycle)-($n$-cycle)-($n$-cycle)-(2-cycle) which lift to arbitrary primaries on the cover, and descendants thereof.  In such cases a single parameter in the map $s$ parameterizes both the base space cross ratio $\zeta_z$ and the covering space cross ratio $\zeta_t$.  We find that the descent relations for the four point functions depend only on base space data and the parameter $s$, which we argue is tantamount to writing the descent relations in terms of the base space data and the base space cross ratio.  These descent relations again do not depend on the covering space data, nor the specifics of the seed CFT, making these universal as well.
\end{abstract}

\end{titlepage}

\tableofcontents

\pagenumbering{arabic}
\section{Introduction}

AdS/CFT \cite{Maldacena:1997re,Witten:1998qj} has become a fundamental tool in theoretical physics, providing insight into quantum gravity, and also providing new tools for modeling strongly coupled systems.  Finding which types of theories admit gravitational limits and connecting them to their gravitational duals remains an open problem, although there is guidance \cite{Heemskerk:2009pn}.  In this work, we concentrate on theories within the context of AdS$_3$/CFT$_2$.

The D1/D5 system is an extremely well studied setup in this class \cite{Strominger:1996sh,Giveon:1998ns,Vafa:1995bm}, and provides an instance with an AdS$_3$ bulk and a known field theory dual (see also \cite{David:2002wn}).  For a special point in the moduli space, one can argue that it is a free orbifold CFT \cite{Dijkgraaf:1998gf,deBoer:1998ip,Seiberg:1999xz,Larsen:1999uk}.  The generality of the orbifolding technique \cite{Dijkgraaf:1989hb} can be used to build large $N$ models, and strongly suggests this as road to other holographic CFTs \cite{Haehl:2014yla,Belin:2014fna,Belin:2015hwa,Belin:2016yll,Belin:2017jli}.  In addition to being candidates as holographic models, they also admit a host of calculational techniques that can be brought to bear \cite{Dijkgraaf:1996xw,Lunin:2000yv,Lunin:2001pw,Pakman:2009ab, Pakman:2009zz, Pakman:2009mi,Keller:2011xi}.  These make orbifold CFTs, particularly when orbifolding by the symmetric group $S_N$, interesting avenues of investigation.

While these models admit large $N$/large central charge limits, one must still confront the deformation away from the orbifold fixed point to the strong coupling regime.  This is where these theories are believed to be well described by gravity \cite{Heemskerk:2009pn,Haehl:2014yla,Belin:2014fna,Belin:2015hwa,Belin:2016yll,Belin:2017jli}.  This has been studied vigorously \cite{Guo:2022ifr,Guo:2022sos,AlvesLima:2022elo,Lima:2021wrz,Lima:2020boh,Lima:2020urq,Keller:2019suk,GarciaiTormo:2018vqv,Hampton:2018ygz,Burrington:2017jhh, Carson:2016uwf,Gaberdiel:2015uca,Carson:2014xwa, Carson:2014ena, Carson:2014yxa,Burrington:2014yia,Burrington:2012yq,Avery:2010vk, Avery:2010er, Avery:2010hs,Avery:2009tu, Avery:2009xr}, often in the context of the D1D5 CFT where the deformation operator is in the twisted sector of the orbifold theory \cite{David:1999ec,Lunin:2001pw}.

However, given that one can start with a generic seed CFT, one may ask question about the ubiquity of certain results.  Indeed, this was our general line of investigation in our earlier work \cite{Burrington:2018upk}, where the universal nature four-point correlators of bare twists at large $N$ suggested a universal nature of the crossing channels.  Indeed, a set of crossing channels was found in the form of fractional modes of the stress tensor acting on bare twists, and these were shown to reproduce all crossing channels to several non-leading orders \cite{Burrington:2018upk}.  A similar structure was seen in the supersymmetric case \cite{DeBeer:2019oxm}.  Other studies also address fractional modes, including \cite{Dei:2019iym,Roumpedakis:2018tdb}, and importantly \cite{Lunin:2001pw} where fractional modes of the superconformal algebra were used to construct the deformation operator of the D1/D5 orbifold CFT by exciting a bare twist.

This prompted our investigation \cite{Burrington:2022dii} to more seriously investigate the fractional modes.  In \cite{Burrington:2022dii}, we considered how covering space techniques \cite{Lunin:2000yv, Lunin:2001pw,Burrington:2012yn} could be used to investigate fractional descendants of fields, and how correlations of these fields could be related to the ancestors in the large $N$ limit, where the covering surface is a sphere.  First, we established how the full Virasoro generators $L$, the fractional virasoro generators $\ell$, and the covering space Virasoro generators ${\mc L}$ can be related through the covering space map.  This allowed us to identify the proper set of ancestors as those operators that lift to primary operators on the cover.  Next, we showed how the mixed commutation relations between full Virasoro generators $L$ and fractinal Virasoro generators $\ell$ close on the fractional Virasoro generators $\ell$. This allowed us to attack the problem in stages, first by commuting all $L$ operators to the left of all $\ell$ operators, which allowed the treatment of the full Virasoro generators $L$ through contour pulls on the base space.  The remaining correlators contained only $\ell$ excitations, and these calculations could be lifted to the covering surface.  Upon lifting, the $\ell$ operators are expanded in terms of covering space Virasoro generators ${\mc L}$.  The important point here is that such expansions necessarily terminate because the expansion on the cover is an {\it operator} valued Laruent series, and this acts on a module with an ancestor of lowest conformal dimension.  Thus, operator coefficients in this expansion ${\mc L}_{n}$ with $n$ too large annihilate the operator, therefore truncating the series on the cover, removing all branch cuts.  These covering surface ladder operators may then be addressed with contour pulls on the cover.  The results from such a procedure seem to involve information from the covering surface and the base space, even though the original problem may be written in terms of base space operators.  We argue that covering space information must cancel.  We tested this by explicit calculation on some example three point functions of simple cycles of the form $(5)-(2)-(4)$ which lift to primaries on the cover.

It is the purpose of this work to extend these results in two directions.  First, in section \ref{3ptsection}, we consider general length single cycle operators in three point functions.  The covering space map for these has been previously worked out \cite{Lunin:2000yv}, and here we use this to explore the fractional descendants.  While more complicated, this is in principle similar to our example calculations in \cite{Burrington:2022dii}, and we again find that the descent relations are universal in nature.  When we say that they are universal, we mean that they do not appear to depend on the details of the seed CFT, nor on the particular operators in that seed CFT that are used to construct the primaries on the cover.  Rather, the descent relations only depend on the central charge of the seed CFT, the total weight of each of the operators in the base, and the size of the cycles.  Next, in section \ref{4ptsection} we consider an infinite class of four point functions with twist structure $(n)-(2)-(2)-(n)$.  In this case, an additional complication arises because there are cross ratios on the base and on the cover, however, both of these are related to a single parameter $s$ in the covering space map.  To further analyze this case, we consider a four-point function in this twist class that lifts to four point function of arbitrary primaries on the cover.  We find that the correlators of fractional descendants are related to ancestors through base space information, and a function of the parameter $s$.  We argue that, because $s$ is directly related to the cross ratio on the base, it is functionally base space information, although a sum over orbifold images may be necessary at that point.  Again, these descent relationships are found to be universal, involving only coarse information about the CFT through the central charge $c$ of the seed CFT, and coarse information about the operators involved including their twist sector, their total conformal weights, and the cross ratio dependent function on the cover used to build the cross ratio dependent function on the base (which is equivalent to knowing the conformal blocks and structure constants relevant for the four point function).  We end with a discussion of our results and future directions in section \ref{discsection}.  To keep the paper as self contained as possible, we include several appendices, the first of which \ref{JacobiSection} gives an account of Jacobi polynomials which appear in the covering space maps of section \ref{3ptsection}.

\section{Single cycle $(m)-(n)-(q)$ fusion }
\label{3ptsection}

Consider the arbitrary fusion of three single twist operators where the lift is a sphere, i.e. the strict large $N$ limit.  The family of maps to the covering surface \cite{Lunin:2000yv} is given as a ratio of polynomials $P_{d_1}(u)/P_{d_2}(u)$ with degree $d_1$ in the numerator and $d_2$ in the denominator ($d_1$ assumed larger than $d_2$).  The leading order of the numerator polynomial is taken to be $n$ such that there is a ramified point at $u=0$ with $n^{\rm th}$ order branch cut, and $d_1-d_2=q$ gives the ramified point at $u=\infty$ with a $q^{\rm th}$ order branch cut.  In addition, the total number of solutions for regular points (i.e. the total number of sheets $s$) must give the covering surface as genus zero by the Riemann-Hurwitz formula, i.e. $s=1/2(n+m+q-1)$.  This must match the degree of the numerator polynomial, and so
\begin{align}
d_1=\frac{1}{2}\left(n+m+q-1\right)\geq 0, \qquad d_2=d_1-q\geq 0, \qquad (d_1>d_2).
\end{align}
To give a fusion of $(n)-(m)\rightarrow (q)$, the number of shared indices for the $(n)$ and $(m)$ cycles must have a specific value, which we label by $k_{m,n}=\frac{1}{2}(n+m+1-q)$, which for brevity we simply denote $k_{m,n}=k$.  However, it is useful to sometimes reintroduce the indices to make the interchange symmetry for the twists clear.  If we take $k$ as the indexed form $k=k_{m,n}$, then it is straightforward to show that $q=(m+n-2k_{m,n}+1)$, $m=(n+q-2k_{n,q}+1)$, and $n=(m+q-2k_{m,q}+1)$ given that $k_{n,q}=n-k_{m,n}+1$ and $k_{m,q}=m-k_{m,n}+1$.  These relationships are $m,n,q$ permutation symmetric, noting that ``overlap'' is a symmetric concept, i.e. $k_{i,j}=k_{j,i}$.  These relationships are the ones appropriate to the group product $(1,2,... n)(n,n-1,n-2,..,n-k+1,n+1,n+2..n+m-k)=(1,2,..,n-k+1,n+1,n+2..,n+m-k)$ (and of course conjugacy equivalent group products).

The explicit form of the map was worked out in \cite{Lunin:2000yv} where the $m^{\rm th}$ order branch point was placed at $u=1$, giving
\begin{align}
w&= u^n \frac{P_{d_1-n}^{(n,-d_1-d_2+n-1)}(1-2u)}{P_{d_2}^{(-n,-d_1-d_2+n-1)}(1-2u)}= u^n\frac{P^{(n,-m)}_{\frac{1}{2}\left(m+q-n-1\right)}(1-2u)}{P^{(-n,-m)}_{\frac{1}{2}\left(n+m-q-1\right)}(1-2u)}
=u^n \frac{P^{(n,-m)}_{m-k}(1-2u)}{P^{(-n,-m)}_{k-1}(1-2u)}\nn \\
&= (-1)^{m-1} u^n \frac{P_{m-k}^{(-m,n)}(2u-1)}{P_{k-1}^{(-m,-n)}(2u-1)}
\end{align}
where $P_\gamma^{(\alpha,\beta)}$ are Jacobi polynomials. The above map has ramified points at $u=0,u=1,u=\infty$ which map to the points $w=0, w=1, w=\infty$ respectively, with ramifications $n-1, m-1, q-1$ respectively.  The order of the Jacobi polynomials may be expressed purely in terms of the overlaps $k_{i,j}$, and in certain circumstances it will be convenient to express all quantities in terms of these, i.e.
\begin{equation}
m=k_{m,n}+k_{m,q}-1, \qquad n=k_{m,n}+k_{n,q}-1, \qquad q= k_{m,q}+k_{n,q}-1
\end{equation}
and so, for example
\begin{equation}
w= u^{[k_{m,n}+k_{n,q}-1]} \frac{P^{([k_{m,n}+k_{n,q}-1],-[k_{m,n}+k_{m,q}-1])}_{k_{m,q}-1}(1-2u)}{P^{(-[k_{m,n}+k_{n,q}-1],-[k_{m,n}+k_{m,q}-1])}_{k_{m,n}-1}(1-2u)}.
\end{equation}
Note that $m,n,q$ must be specified with restrictions, e.g. for given $m,n$, the restriction on $q$ is given by $m+n-1\geq q \geq |m-n|+1$ and $m+n+q$ is odd. The $k_{i,j}$, on the other hand, may be independently specified for any integers with $k_{i,j}\geq 1$, making these somewhat more natural.

To make headway with computations, we find the derivative of the above function, which can be expressed using a Wronskian \cite{Lunin:2000yv}, giving
\begin{equation}
\pa_{u} w= \frac{1}{\left(P^{(-m,-n)}_{k-1}(2u-1)\right)^2} u^{n-1}(u-1)^{m-1} \frac{(-1)^{m-k}(n+m-k)!}{(k-1)!(m-k)!(n-k)!}.
\label{dwdugen}\end{equation}
We provide a direct proof of the above in appendix \ref{wronskandschwarz}.

We map the ramified points to finite points in the $t$ and $z$ planes via the $sl(2)$ transformations
\begin{align}
& u(t)=\frac{t_m-t_{q}}{t_m-t_n}\frac{t-t_n}{t-t_{q}} \label{utmnq}\\
& z(w)=\frac{z_{q}(z_m-z_{n})w-z_n(z_m-z_{q})} {(z_m-z_{n})w-(z_m-z_{q})}.
\end{align}
The resulting map, $z(w(u(t)))$ can be shown to be explicitly $m,n,q$ permutation symmetric, which we show in the appendix \ref{mnqsym}.  Using the above map, we can compute the Schwarzian and find that
\begin{align}
\left\{z,t\right\}
&=\left(\left\{z,w\right\}w'^2+\left\{w,u\right\}\right)u'^2+\left\{u,t\right\} \nn\\
&=\left\{w,u\right\}u'^2 \nn \\
&=-\frac{\hat{h}_m^\uparrow}{2(t-t_m)^2} -\frac{\hat{h}_n^\uparrow}{2(t-t_n)^2} -\frac{\hat{h}_{q}^\uparrow}{2(t-t_{q})^2} \nn \\
& \qquad  +\frac{\hat{h}_{m}^\uparrow + \hat{h}_{q}^\uparrow-\hat{h}_{n}^\uparrow}{2(t-t_m)(t-t_{q})}  +\frac{\hat{h}_{m}^\uparrow+\hat{h}_{n}^\uparrow-\hat{h}_{q}^\uparrow}{2(t-t_n)(t-t_m)} +\frac{\hat{h}_{n}^\uparrow+\hat{h}_{q}^\uparrow-\hat{h}_{m}^\uparrow}{2(t-t_n)(t-t_{q})} \label{schwarzgen}
\end{align}
where we use the notation $\hat{h}^\uparrow_{n_i}=n_i^2-1$: this is the ``stripped weight'' of the bare twist $h_{\rm bare}=\frac{c}{24 n_i} \hat{h}^{\uparrow}$.  We prove the above form for the Schwarzian in appendix \ref{wronskandschwarz}.  The Schwarzian is easily seen to be $m,n,q$ permutation symmetric, as is required by the results of appendix \ref{mnqsym}.

We are now in a position to generalize the computations from \cite{Burrington:2022dii} to arbitrary $(m)-(n)-(q)$ fusions.  We will start with the parent three point function
\begin{align}
&\langle \left[\mathcal{O}_{m,{-a_m/m}}\sigma_{m}\right](z_m) \left[\mathcal{O}_{n,{-a_n/n}}\sigma_{n}\right](z_n) \left[\mathcal{O}_{q,{-a_{q}/(q)}}\sigma_{q}\right](z_{q})\rangle
\end{align}
(where we have suppressed the antiholomorphic side, as usual).  In the above, the operators $\mathcal{O}_{n_i}$ are primaries with weight $a_{n_i}$, and the modes $\mathcal{O}_{n_i,-a_{n_i}/n_i}$ are constructed to lift to primaries on the cover (argued in our \cite{Burrington:2022dii} to be the appropriate ``parent'' correlators to explore).  We will need the expansions of the map near the various ramified points in the map $(t=t_m, u=1, w=1,z=z_m), (t=t_n, u=0, w=0, z=z_n), (t=t_q, u=\infty, w=\infty, z=z_{\infty})$.  We first expand in $w$, to find
\begin{align}
& z-z_m = \left(\frac{z_{mq}z_{mn}}{z_{nq}} (w-1)\right) \frac{1}{1-\frac{z_{mn}}{z_{nq}}(w-1)}= \sum_{i=1}^\infty z_{mq}\left(\frac{z_{mn}}{z_{nq}}\right)^i (w-1)^i  \nn \\
& z-z_n = \left(\frac{z_{nq}z_{mn}}{z_{mq}} w\right) \frac{1}{1-\frac{z_{mn}}{{z_{mq}}}w}= \sum_{i=1}^\infty z_{nq}\left(\frac{z_{mn}}{z_{mq}}\right)^i w^i \label{zexpandw}\\
& z-z_q = -\left(\frac{z_{mq}z_{nq}}{z_{mn}}\frac{1}{w}\right)\frac{1}{1-\frac{z_{mq}}{z_{mn}}\frac{1}{w}}=\sum_{i=1}^\infty z_{nq}\left(\frac{z_{mq}}{z_{mn}}\right)^i \frac{1}{w^i}. \nn
\end{align}
Let us consider the expansion around $w=0$, and the subsequent expansions in $w(u(t))$ near $t=t_n$.  Due to the map $w(u)$ being ramified at the point $u=0$, the first $n-1$ non-trivial terms in the expansion of $w(u(t))$ near $t=t_n$ are contained in the linear term of the expansion of $z(w)$, i.e. $\frac{z_{mn}z_{nq}}{z_{mq}}w$.  Thus, if we wish to consider modes $\ell_{-b/n}$ with $b<n$, this first term in the expansion is all that will be necessary.  This can be generalized to the statement that for $\ell_{-b/n}$ with $(r-1)n \leq b < rn$ ($r$ some integer) may truncate to order $w^r$ terms in the expansion of $z(w)$ when operating on an operator that lifts to a primary on the cover.  Furthermore, the $m,n,q$ interchange symmetry of appendix \ref{mnqsym} can be exploited to cut down on repetitive calculations.

To move forward, we lift the operator at location $z_n$ to the cover
\begin{equation}
\left[\mathcal{O}_{n,{-a_n/n}}\sigma_{n}\right](z_n) \rightarrow \oint_{t_n} \frac{dt}{2\pi i} (z(t)-z_n)^{a_n - a_n/n-1} (\pa z)^{-a_n+1} \mathcal{O}_n(t).
\end{equation}
The map is ramified as $z-z_n = A_n (t-t_n)^n +\cdots$, and we can see that the leading order term produces the simple pole in $t-t_n$ in the measure of the contour integral.  Thus, the leading order contribution depends only on the constant $A_n$ through
\begin{equation}
\left[\mathcal{O}_{n,{-a_n/n}}\sigma_{n}\right](z_n) \rightarrow n^{-a_n+1} A_{n}^{-a_n/n}\mathcal{O}_n(t_n).
\end{equation}
To extract $A_n$, we simply plug in $w(u)$ into $z-z_n$ finding
\begin{equation}
z-z_n = \left(\frac{z_{nq}z_{mn}}{z_{mq}} (-1)^{m-1}u^{n}\frac{P^{-m,n}_{m-k}}{P^{-m,-n}_{k-1}}\right) \frac{1}{1-\frac{z_{mn}}{{z_{mq}}}(-1)^{m-1}u^{n}\frac{P^{-m,n}_{m-k}}{P^{-m,-n}_{k-1}}}.
\end{equation}
Because $u$ goes to $0$ linearly as $t$ goes to $t_n$, the right most term may be set to $1$
\begin{equation}
\frac{1}{1-\frac{z_{mn}}{{z_{mq}}}(-1)^{m-1}u^{n}\frac{P^{-m,n}_{m-k}}{P^{-m,-n}_{k-1}}}\rightarrow 1.
\end{equation}
This remains true for additional excitations involving fractional virasoro generators, so long as the index of $\ell_{-b/n}$ has $b<n$.  Thus, substituting in $u(t)\approx \frac{t_{mq}}{t_{mn}t_{nq}}(t-t_n)$ into $u^n$ and $u=0$ elsewhere, we find
\begin{align}
z-z_n & = \frac{z_{nq}z_{mn}}{z_{mq}}(-1)^{m-1}\left(\frac{t_{mq}}{t_{nq}t_{mn}}\right)^n \frac{P^{-m,n}_{m-k}(-1)}{P^{-m,-n}_{k-1}(-1)}(t-t_n)^n + \cdots \nn \\
\end{align}
allowing us to identify
\begin{align}
A_n &= \frac{z_{nq}z_{mn}}{z_{mq}}(-1)^{m-1}\left(\frac{t_{mq}}{t_{nq}t_{mn}}\right)^n \frac{P^{-m,n}_{m-k}(-1)}{P^{-m,-n}_{k-1}(-1)} \nn \\
&= \frac{z_{mn}z_{nq}}{z_{mq}}\left(\frac{t_{mq}}{t_{nq}t_{mn}}\right)^n \frac{(-1)^{k-1}(n+m-k)!(k-1)!(n-k)!}{n((n-1)!)^2(m-k)!} \\
&=  \frac{z_{mn}z_{nq}}{z_{mq}} \left(\frac{t_{mq}}{t_{nq}t_{mn}}\right)^n \frac{(-1)^{\frac{m+n-q-1}{2}}\left(\frac{m+n+q-1}{2}\right)!\left(\frac{m+n-q-1}{2}\right)! \left(\frac{n+q-m-1}{2}\right)!}{n((n-1)!)^2\left(\frac{m+q-n-1}{2}\right)!} \\
&\equiv \frac{z_{mn}z_{nq}}{z_{mq}}\left(\frac{t_{mq}}{t_{nq}t_{mn}}\right)^n B_n \label{defBn}
\end{align}
where the particular value of the Jacobi polynomials has been read using (\ref{valueat1}) and (\ref{symu}).  Furthermore, we have pulled out the numeric coefficient, and defined this as $B_n$.

The operators at other locations are similarly found, either using the Jacobi polynomial identities in giving appendix \ref{mnqsym}, or simply using the $m,n,q$ interchange symmetry directly, finding
\begin{align}
\left[\mathcal{O}_{m,{-a_m/m}}\sigma_{m}\right](z_m) \rightarrow m^{-a_m+1} A_{m}^{-a_m/m}\mathcal{O}_m(t_m) \\
\left[\mathcal{O}_{q,{-a_q/q}}\sigma_{q}\right](z_q) \rightarrow q^{-a_q+1} A_{q}^{-a_q/q}\mathcal{O}_q(t_q)
\end{align}
with
\begin{align}
A_m &= \frac{z_{mn}z_{mq}}{z_{nq}} \left(\frac{t_{nq}}{t_{mq}t_{mn}}\right)^m \frac{(-1)^{\frac{m+q-n-1}{2}}\left(\frac{m+n+q-1}{2}\right)!\left(\frac{m+n-q-1}{2}\right)! \left(\frac{m+q-n-1}{2}\right)!}{m((m-1)!)^2\left(\frac{n+q-m-1}{2}\right)!} \\
&\equiv \frac{z_{mn}z_{mq}}{z_{nq}} \left(\frac{t_{nq}}{t_{mq}t_{mn}}\right)^m  B_m \\
A_q &= \frac{z_{nq}z_{mq}}{z_{mn}} \left(\frac{t_{mn}}{t_{nq}t_{mq}}\right)^q \frac{(-1)^{\frac{q+n-m-1}{2}}\left(\frac{m+n+q-1}{2}\right)!\left(\frac{m+q-n-1}{2}\right)! \left(\frac{n+q-m-1}{2}\right)!}{q((q-1)!)^2\left(\frac{m+n-q-1}{2}\right)!} \\
&\equiv \frac{z_{nq}z_{mq}}{z_{mn}} \left(\frac{t_{mn}}{t_{nq}t_{mq}}\right)^q   B_q. \\
\end{align}
The most efficient way to relate the coefficients $A_i$ is to perform a cyclic $m\rightarrow n\rightarrow q\rightarrow m$ interchange, under which the ratios involving $t_{ij}$ and $z_{ij}$ transform into each other without any extra minus signs needed, leading to a direct relationship between the $B_i$ under such an interchange.

Finally, the resulting three point function is
\begin{align}
&\langle \left[\mathcal{O}_{m,{-a_m/m}}\sigma_{m}\right](z_m) \left[\mathcal{O}_{n,{-a_n/n}}\sigma_{n}\right](z_n) \left[\mathcal{O}_{q,{-a_{q}/(q)}}\sigma_{q}\right](z_{q})\rangle\nn \\
&\rightarrow   \left(\frac{z_{mn}z_{mq}}{z_{nq}}\right)^{-a_m/m} \left(\frac{z_{mn}z_{nq}}{z_{mq}}\right)^{-a_n/n} \left(\frac{z_{mq}z_{nq}}{z_{mn}}\right)^{-a_q/q} \nn \\
& \qquad \times m^{-a_m+1}n^{-a_n+1}q^{-a_q+1} B_{m}^{-a_m/m}B_{n}^{-a_n/n}B_{q}^{-a_q/q} C^{\uparrow}_{\mathcal{O}_m,\mathcal{O}_n,\mathcal{O}_q} \label{barefuncform}
\end{align}
where $C^{\uparrow}_{\mathcal{O}_m,\mathcal{O}_n,\mathcal{O}_q}$ is the covering space structure constant for the three point function on the cover.  We have assumed that the original operators in question $\mathcal{O}_i$ on the base space were normalized, and so their lifts to the cover are as well.  We note that the powers of $t_{ij}$ appearing in the $A_i$ are just such that they cancel the three point function dependence on the cover, and so no functional dependence on the $t_{ij}$ remain.

Multiplying the above (\ref{barefuncform}) by the Liouville term, we find that the remaining powers of $z_{ij}$ above adjusts the three point function to that of a three point function of primaries with conformal weights
\begin{align}
& h_{m}=\frac{c}{24}(m-1/m) + a_m/m, \nn \\
& h_{n}=\frac{c}{24}(n-1/n) + a_n/n, \\
& h_{q}=\frac{c}{24}(q-1/q) + a_q/q \nn
\end{align}
as expected.  One may normalize by the two point functions using the maps $z(t)=t^i/((1-t)^i+t^i)$ for the twist $i$ operators.  While this is straightforward, this is not our main purpose.  We wish to construct the correlators of fractional conformal descendants, given the correlator of the ancestors.  Thus, we absorb all of the numerical constants into a composite structure constant, writing
\begin{equation}
\langle \left[\mathcal{O}_{m,{-a_m/m}}\sigma_{m}\right](z_m) \left[\mathcal{O}_{n,{-a_n/n}}\sigma_{n}\right](z_n) \left[\mathcal{O}_{q,{-a_{q}/(q)}}\sigma_{q}\right](z_{q})\rangle = \frac{C_{\mathcal{O}_{m,{-a_m/m}},\mathcal{O}_{n,{-a_n/n}},\mathcal{O}_{q,{-a_{q}/q}}}} {z_{mn}^{h_m+h_n-h_q}z_{mq}^{h_m+h_q-h_n}z_{nq}^{h_n+h_q-h_m}}.
\end{equation}
While this is in some sense a trivial statement about the excited twist field being primaries, the above calculation fleshes out how this happens from the covering space calculation, and will allow us to identify factors that appear in common when calculating the correlators of fractional conformal descendants.

We now consider the lowest weight descendants
\begin{align}
&\langle \left[\ell_{-1/m}\mathcal{O}_{m,{-a_m/m}}\sigma_{m}\right](z_m) \left[\mathcal{O}_{n,{-a_n/n}}\sigma_{n}\right](z_n) \left[\mathcal{O}_{q,{-a_{q}/(q)}}\sigma_{q}\right](z_{q})\rangle, \\
&\langle \left[\mathcal{O}_{m,{-a_m/m}}\sigma_{m}\right](z_m) \left[\ell_{-1/n}\mathcal{O}_{n,{-a_n/n}}\sigma_{n}\right](z_n) \left[\mathcal{O}_{q,{-a_{q}/(q)}}\sigma_{q}\right](z_{q})\rangle, \\
&\langle \left[\mathcal{O}_{m,{-a_m/m}}\sigma_{m}\right](z_m) \left[\mathcal{O}_{n,{-a_n/n}}\sigma_{n}\right](z_n) \left[\ell_{-1/q}\mathcal{O}_{q,{-a_{q}/(q)}}\sigma_{q}\right](z_{q})\rangle.
\end{align}
However, the three above computations should be related by the $m,n,q$ permutation symmetry.  We will explicitly compute for the case where the excitation acts on the operator at location $z_n$, reading the others by exchanging $m,n,q$ appropriately.

First we lift
\begin{align}
&\langle \left[\mathcal{O}_{m,{-a_m/m}}\sigma_{m}\right](z_m) \left[\ell_{-1/n}\mathcal{O}_{n,{-a_n/n}}\sigma_{n}\right](z_n) \left[\mathcal{O}_{q,{-a_{q}/(q)}}\sigma_{q}\right](z_{q})\rangle \nn \\
& \rightarrow m^{-a_m+1} A_{m}^{-a_m/m}n^{-a_n+1} A_{n}^{-a_n/n}q^{-a_q+1} A_{q}^{-a_q/q} \nn \\
& \qquad \times
\langle\mathcal{O}_{m}(t_m) \left[\oint_{t_n}\frac{dt}{2\pi i} (z(t)-z_{n})^{-\frac{1}{n}+1}(\pa z)^{-1}\left(T(t)-\frac{c}{12}\left\{z(t),t\right\}\right)\mathcal{O}_{n}(t_n) \right] \mathcal{O}_{q}(t_q)  \rangle. \label{liftLm1ovn}
\end{align}
The term $z(t)-z_n$ goes to 0 as $w^1 \sim u^n \sim (t-t_n)^n$, and $(\pa z)$ goes to 0 as $(t-t_n)^{(n-1)}$ and so the measure $(z(t)-z_n)^{(-1/n+1)}(\pa z)^{-1}$ goes to a constant.  Given that the operator $\mathcal{O}_{n}$ is primary on the cover, we must only expand the measure by a single term (this is true for the Schwarzian term as well, because it goes to infinity as $1/(t-t_n)^2$).  Thus, we must expand the measure once beyond the leading term.  Thus, so long as $n\geq 2$, we can use the leading order $w$ expansion of (\ref{zexpandw}).  This is generally true, given an operator $\ell_{-f/n}$ with $f<n$ we only need the leading order term in the $w$ expansion of (\ref{zexpandw}) when acting on a primary on the cover.  We find, using only this first term
\begin{align}
(z(t)-z_n) & \approx \frac{z_{mn}z_{nq}}{z_{mq}} u^n \frac{P^{(n,-m)}_{\frac{1}{2}(m-n+q-1)}(1-2u)}{P^{(-n,-m)}_{\frac{1}{2}(m+n-q-1)}(1-2u)}
\end{align}
where the above is to be read as $u=u(t)$.  Expanding $u(t)$ is relatively straightforward.  However, the ratio of Jacobi polynomials may seem at first problematic, given that the degrees of the polynomials vary, depending on the particular $m,n,q$ at hand.  However, we note that reexpressing these in terms of the overlaps $k_{i,j}$, we find
\begin{align}
\frac{P^{(n,-m)}_{\frac{1}{2}(m-n+q-1)}(1-2u)}{P^{(-n,-m)}_{\frac{1}{2}(m+n-q-1)}(1-2u)}
&=\frac{P^{([k_{m,n}+k_{n,q}-1],-[k_{m,n}+k_{m,q}-1])}_{k_{m,q}-1}(1-2u)} {P^{(-[k_{m,n}+k_{n,q}-1],-[k_{m,n}+k_{m,q}-1])}_{k_{m,n}-1}(1-2u)} \nn \\
&= \frac{\sum_{\ell=0}^{k_{m,q}-1} \frac{(k_{n,q})_{\ell}(k_{m,n}+k_{n,q}+\ell)_{k_{m,q}-1-\ell}}{\ell!(k_{m,q}-1-\ell)!}(-u)^\ell}{\sum_{\ell=0}^{k_{m,n}-1} \frac{(-k_{m,n}-k_{n,q}-k_{m,q}+2)_\ell(-k_{m,n}-k_{n,q}+2+\ell)_{k_{m,n}-1-\ell}}{\ell!(k_{m,n}-1-\ell)!}(-u)^\ell} \nn \\
&= \frac{\sum_{\ell=0}^{k_{m,q}-1} \frac{(k_{n,q})_{\ell}(k_{m,n}+k_{n,q}+\ell)_{k_{m,q}-1-\ell}}{\ell!(k_{m,q}-1-\ell)!}(-u)^\ell} {(-1)^{k_{m,n}-1}\sum_{\ell=0}^{k_{m,n}-1} \frac{(k_{m,n}+k_{n,q}+k_{m,q}-1-\ell)_\ell(k_{n,q})_{k_{m,n}-1-\ell}}{\ell!(k_{m,n}-1-\ell)!}(-u)^\ell}
\end{align}
We note that the numerator has no problematic terms in the pochhammer symbols: all terms are positive definite in the products.  The denominator always starts with a finite term because the pochammer symbols for $\ell=0$ are non-zero.  Any further zeros in the pochammer symbols, thus, do not cause problems (and in fact, one can show that all of the pochammer symbols are non-zero for the ranges of allowed $\ell$).

However, we wish to extend the range of the sums to obtain the most universal answers we can.  Recall that the above expression is only useful in the case where we are exploring a term of order $\ell_{-p/n}$ where $0\leq p \leq n-1$.  Thus, most optimistically, we would like to take both sums to have upper bounds $\ell_{\rm max}=n-1$.  First we address the numerator
\begin{equation}
P^{([k_{m,n}+k_{n,q}-1],-[k_{m,n}+k_{m,q}-1])}_{k_{m,q}-1}(1-2u)={\sum_{\ell=0}^{k_{m,q}-1} \frac{(k_{n,q})_{\ell}(k_{m,n}+k_{n,q}+\ell)_{k_{m,q}-1-\ell}}{\ell!(k_{m,q}-1-\ell)!}(-u)^\ell} \end{equation}
Note that the pochammer symbols above are always products of positive numbers for the ranges given.  To extend the sum, we regulate the terms in the summand by replacing $\ell\rightarrow \ell-\epsilon$ and taking the limit as $\epsilon\rightarrow 0$.  For $k_{m,q}-1<\ell$, the denominator of the sum contains $(k_{m,q}-1-\ell+\epsilon)!$, which we interpret through a gamma function regularization $1/[(k_{m,q}-1-\ell)!]=1/[\lim_{\epsilon\rightarrow 0}\Gamma(k_{m,q}-\ell+\epsilon)]=0$ making this appear to be a ``regulating'' term, cutting off the sum where appropriate.  However, we must also contend with pochhammer symbols of negative subscript.   If we allow for pochhammer symbols with negative subscript, defining them by the relation $(\kappa)_{\delta}=\Gamma(\kappa+\delta)/\Gamma(\kappa)$, we can see that the only singular cases are when $\kappa+\delta\leq 0$ and $\kappa \geq 1$: these conditions never happen in the numerator.  Thus, we may safely replace the upper bound of the sum in the numerator by $\infty$, or, as we have mentioned, the more proper $n-1$.

Dealing with the denominator, we see that the sum
\begin{align}
& P^{(-[k_{m,n}+k_{n,q}-1],-[k_{m,n}+k_{m,q}-1])}_{k_{m,n}-1}(1-2u) \nn \\
& \qquad = {(-1)^{k_{m,n}-1}\sum_{\ell=0}^{k_{m,n}-1} \frac{(k_{m,n}+k_{n,q}+k_{m,q}-1-\ell)_\ell(k_{n,q})_{k_{m,n}-1-\ell}} {\ell!(k_{m,n}-1-\ell)!}(-u)^\ell}
\end{align}
has upper bound $k_{m,n}-1$ which we would also like to extend. Again, we use the shift $\ell\rightarrow \ell-\epsilon$ in the summand and relax $\epsilon$ to 0 to define any singular terms.  Note that, as before,  $1/[(k_{m,n}-1-\ell)!]=1/[\lim_{\epsilon\rightarrow 0}\Gamma(k_{m,n}-\ell+\epsilon)]=0$ acts as a regulating term.  However, the pochhammer symbol $(k_{n,q})_{k_{m,n}-1-\ell+\epsilon}=\Gamma(k_{m,n}+k_{n,q}-1-\ell+\epsilon)/\Gamma(k_{n,q})$ becomes singular as $\epsilon\rightarrow 0$ for sufficiently large $\ell$.  This happens when
\begin{equation}
k_{m,n}+k_{n,q}-1-\ell=n-\ell \leq 0
\end{equation}
Thus, so long as we only expand the denominator to $\ell_{\rm max}=n-1$, the factorial in the denominator $1/[(k_{m,n}-1-\ell)!]$ acts as a regulator to limit the sum. \footnote{Interestingly, the same regularization of all terms in the sum in the denominator leads to a finite window $k_{m,n}+k_{n,q}-1\leq \ell \leq k_{m,n}+k_{n,q}+k_{m,q}-2$ where the sum can also be evaluated (in addition to $0\leq \ell \leq k_{m,n}-1$).  Summing over this distinct window and regulating we find
\begin{align}
&{(-1)^{k_{m,n}-1}\sum_{\ell=k_{m,n}+k_{n,q}-1}^{k_{m,n}+k_{n,q}+k_{m,q}-2} \frac{(k_{m,n}+k_{n,q}+k_{m,q}-1-\ell+\epsilon)_{\ell-\epsilon}(k_{n,q})_{k_{m,n}-1-\ell+\epsilon}} {(\ell-\epsilon)!(k_{m,n}-1-\ell+\epsilon)!}(-u)^\ell} \nn \\
=& {(-1)^{k_{m,n}-1}(-u)^{k_{m,n}+k_{n,q}-1}\sum_{\ell=0}^{k_{m,q}-1} \frac{(k_{m,q}-\ell+\epsilon)_{k_{m,n}+k_{n,q}-1+\ell-\epsilon}(k_{n,q})_{-k_{n,q}-\ell+\epsilon}} {(k_{m,n}+k_{n,q}-1+\ell-\epsilon)!(-k_{n,q}-\ell+\epsilon)!}(-u)^\ell} \nn \\
=& -(u)^n \sum_{\ell=0}^{k_{m,q}-1} \frac{(k_{n,q})_{\ell}(k_{m,n}+k_{n,q}+\ell)_{k_{m,q}-1-\ell}}{\ell!(k_{m,q}-1-\ell)!}(-u)^\ell \nn \\
=& -(u)^n P^{([k_{m,n}+k_{n,q}-1],-[k_{m,n}+k_{m,q}-1])}_{k_{m,q}-1}(1-2u)
\end{align}
where we have used $\Gamma(-A+\epsilon)/\Gamma(-A-\delta+\epsilon)=(-1)^{\delta}\Gamma(A+\delta+1-\epsilon)/\Gamma(A+1-\epsilon)$ where $\delta$ is an integer, and if both $A$ and $\delta$ are integers, the equality holds exactly in the limit as $\epsilon\rightarrow 0$.  This effectively unites the numerator and denominator as two parts of the same sum.}

Thus, for purposes of expansion to an order $u^n\mathcal{O}(u^p)$ with $p\leq n-1$, we may use
\begin{align}
(z(t)-z_n) & \approx \frac{z_{mn}z_{nq}}{z_{mq}} u^n\frac{\sum_{\ell=0}^{n-1} \frac{(k_{n,q})_{\ell}(k_{m,n}+k_{n,q}+\ell)_{k_{m,q}-1-\ell}}{\ell!(k_{m,q}-1-\ell)!}(-u)^\ell} {(-1)^{k_{m,n}-1}\sum_{\ell=0}^{n-1} \frac{(k_{m,n}+k_{n,q}+k_{m,q}-1-\ell)_\ell(k_{n,q})_{k_{m,n}-1-\ell}}{\ell!(k_{m,n}-1-\ell)!}(-u)^\ell}
\end{align}
Doing so, we find
\begin{align}
(z(t)-z_n) & \approx \frac{z_{mn}z_{nq}}{z_{mq}}B_n u^n \bigg(1+\frac{n(m^2+n^2-q^2-1)}{2(n-1)(n+1)}u \nn \\
& \qquad +\frac{Q_{n,2}(m,n,q)}{16(n+1)(n-1)^2(n-2)}u^2+\frac{Q_{n,3}(m,n,q)}{96(n+3)(n-1)^3(n-2)(n-3)}u^3+\cdots\bigg)
\end{align}
where
\begin{align}
B_n &= \frac{(-1)^{\frac{m+n-q-1}{2}}\left(\frac{m+n+q-1}{2}\right)!\left(\frac{m+n-q-1}{2}\right)! \left(\frac{n+q-m-1}{2}\right)!}{n((n-1)!)^2\left(\frac{m+q-n-1}{2}\right)!}
\end{align}
as in (\ref{defBn}), and the rather opaque polynomials
\begin{align}
Q_{n,2}(m,n,q)=&n\big(2m^4n+4m^2n^3-4m^2nq^2 +2n^5-4n^3q^2+2nq^4 \nn \\
& \qquad \qquad -5m^4-2m^2n^2+10m^2q^2-n^4+6n^2q^2-5q^4 -20m^2n-12n^3+12nq^2 \nn \\ & \qquad \qquad \qquad +18m^2+6n^2-14q^2+18n-13
\big)u^2+\cdots\bigg) \\
Q_{n,3}(m,n,q)& =n(2m^6n^2+6m^4n^4-6m^4n^2q^2 \nn \\
&+6m^2n^6-12m^2n^4q^2+6m^2n^2q^4+2n^8-6n^6q^2 \nn\\ &+6n^4q^4-2n^2q^6-13m^6n-15m^4n^3+39m^4nq^2-3m^2n^5 \nn \\ &+42m^2n^3q^2-39m^2nq^4-n^7+15n^5q^2-27n^3q^4 \nn \\ &+13nq^6+23m^6-73m^4n^2-69m^4q^2-99m^2n^4+78m^2n^2q^2 \\ &+69m^2q^4-35n^6+63n^4q^2-5n^2q^4-23q^6+239m^4n+90m^2n^3 \nn\\ &-378m^2nq^2+23n^5-162n^3q^2+139nq^4-157m^4+440m^2n^2+270m^2q^2\nn \\ &+193n^4-128n^2q^2-113q^4-775m^2n \nn \\
&-187n^3+451nq^2+341m^2-337n^2-233q^2+549n-207)\nn
\end{align}
One can immediately see the structure: to believe the $u^3$ term, it must be that $n-1\geq 3$ and so, $n\geq 4$, making the denominator of this term well defined.  Interestingly, these terms are all well defined, specifying $m,q$ first (which restricts possible values of $n$).  For example, if $m=q$, then $n$ can be 1.  Plugging in first $q=m$, we see that
\begin{align}
(z(t)-z_n)& \approx \frac{z_{mn}z_{nq}}{z_{mq}}B_n u^n\bigg(1+\frac{1}{2}nu +\frac{n(2n^3+4m^2+3n^2-8n-13)}{16(n^2-4)}u^2+\nn \\
&\qquad + \frac{n(2n^3+12m^2+5n^2-8n-23)}{96(n-2)}u^3+\cdots\bigg),
\end{align}
but in such a case $k_{n,q}=k_{m,n}$ and so $n=2k_{m,n}-1$ is odd, rendering all terms finite (although they are not to be trusted beyond the appropriate $n$; they begin to interfere with the higher powers of $w$ appearing in the expansion of $z(w)$).

With this expression, it is straightforward to expand to several orders in $t-t_n$, giving
\begin{align}
&(z(t)-z_n) \approx \frac{z_{mn}z_{nq}}{z_{mq}}B_n \left(\frac{t_{mq}}{t_{mn}t_{nq}}\right)^n (t-t_n)^n  \nn \\
& \times\bigg(1+\frac{n((-(n^2+m^2-q^2-1)t_q-(n^2+q^2-m-1)t_m+2(n^2-1)t_n)}{2(n^2-1)t_{mn}t_{nq}}(t-t_n) \nn \\
& \qquad +\frac{Q_{t,n,2}(m,n,q)}{16t_{mn}^2t_{nq}^2(n-1)^2(n^2-4)}(t-t_n)^2 +\frac{Q_{t,n,3}(m,n,q)}{96t_{mn}^3t_{nq}^3(n-1)^3(n-2)(n^2-9)}(t-t_n)^3+\cdots\bigg)
\end{align}
with $Q_{t,n,2}(m,n,q)$ and $Q_{t,n,3}(m,n,q)$ polynomials in $m,n,q$ and $t_m, t_n, t_q$ ($Q_{t,n,2}(m,n,q)$ is displayed in appendix \ref{appxpoly}).  This expansion also gives
\begin{align}
&\pa_t z(t) \approx \frac{z_{mn}z_{nq}}{z_{mq}}B_n \left(\frac{t_{mq}}{t_{mn}t_{nq}}\right)^n n (t-t_n)^{n-1}  \nn \\
& \times\bigg(1+\frac{n+1}{n}\frac{n((-(n^2+m^2-q^2-1)t_q-(n^2+q^2-m-1)t_m+2(n^2-1)t_n)}{2(n^2-1)t_{mn}t_{nq}}(t-t_n) \nn \\
& \qquad +\frac{n+2}{n}\frac{Q_{t,n,2}(m,n,q)}{16t_{mn}^2t_{nq}^2(n-1)^2(n^2-4)}(t-t_n)^2 \nn \\
&\qquad +\frac{n+3}{n}\frac{Q_{t,n,3}(m,n,q)}{96t_{mn}^3t_{nq}^3(n-1)^3(n-2)(n^2-9)}(t-t_n)^3+\cdots\bigg)
\end{align}
We are finally in a position to evaluate the expansion (\ref{liftLm1ovn}).  It is straightforward to check that the Schwarzian term does not contribute (this is equivalent to the statement that $\ell_{-1/n}$ is 0 when acting on a bare twist).  The other terms become
\begin{align}
&m^{-a_m+1} A_{m}^{-a_m/m}n^{-a_n+1} A_{n}^{-a_n/n}q^{-a_q+1} A_{q}^{-a_q/q} \nn \\
& \times \langle\mathcal{O}_{m}(t_m) \left[\oint_{t_n}\frac{dt}{2\pi i} (z(t)-z_{n})^{-\frac{1}{n}+1}(\pa z)^{-1}\left(T(t)\right)\mathcal{O}_{n}(t_n) \right] \mathcal{O}_{q}(t_q)  \rangle \nn \\
&= m^{-a_m+1} A_{m}^{-a_m/m}n^{-a_n+1} A_{n}^{-a_n/n}q^{-a_q+1} A_{q}^{-a_q/q} \nn \\\
&\times \left(\frac{z_{mn}z_{nq}}{z_{mq}}\right)^{\frac{-1}{n}}B_n^{-\frac{1}{n}}\frac{t_{mn} t_{nq}}{t_{mq}}\frac{1}{n} \nn \\
& \langle\mathcal{O}_{m}(t_m) \left[\left({\mc L}_{-1}-2\frac{(m^2-q^2)t_{mq}+(n^2-1)(2 t_n - t_q -t_m)}{(n^2-1)t_{mn}t_{nq}} {\mc L}_0\right)\mathcal{O}_{n}\right](t_n) \mathcal{O}_{q}(t_q)  \rangle
\end{align}
As in our \cite{Burrington:2022dii}, we can make substitutions
\begin{equation}
{\mc L}_{-1} \rightarrow -\frac{-a_m-a_n+a_q}{t_{mn}}-\frac{a_n+a_q-a_m}{t_{nq}}, \qquad {\mc L}_0\rightarrow a_n \label{subsinL0Lm1atn}
\end{equation}
which gives
\begin{align}
&m^{-a_m+1} A_{m}^{-a_m/m}n^{-a_n+1} A_{n}^{-a_n/n}q^{-a_q+1} A_{q}^{-a_q/q} \nn \\
& \times \langle\mathcal{O}_{m}(t_m) \left[\oint_{t_n}\frac{dt}{2\pi i} (z(t)-z_{n})^{-\frac{1}{n}+1}(\pa z)^{-1}\left(T(t)\right)\mathcal{O}_{n}(t_n) \right] \mathcal{O}_{q}(t_q)  \rangle \nn \\
&= m^{-a_m+1} A_{m}^{-a_m/m}n^{-a_n+1} A_{n}^{-a_n/n}q^{-a_q+1} A_{q}^{-a_q/q} \nn \\\
&\times \left(\frac{z_{mn}z_{nq}}{z_{mq}}\right)^{\frac{-1}{n}}B_n^{\frac{-1}{n}}\frac{1}{n}\left(\frac{(q^2-m^2)a_n}{n^2-1}+a_m-a_q\right) \nn \\
& \langle\mathcal{O}_{m}(t_m) \mathcal{O}_{n}(t_n) \mathcal{O}_{q}(t_q)  \rangle
\end{align}
giving that the descendant and ancestor correlation functions are related by
\begin{align}
& \langle \left[\mathcal{O}_{m,{-a_m/m}}\sigma_{m}\right](z_m) \left[\ell_{-1/n}\mathcal{O}_{n,{-a_n/n}}\sigma_{n}\right](z_n) \left[\mathcal{O}_{q,{-a_{q}/(q)}}\sigma_{q}\right](z_{q})\rangle \nn \\
&=\langle \left[\mathcal{O}_{m,{-a_m/m}}\sigma_{m}\right](z_m) \left[\mathcal{O}_{n,{-a_n/n}}\sigma_{n}\right](z_n) \left[\mathcal{O}_{q,{-a_{q}/(q)}}\sigma_{q}\right](z_{q})\rangle \nn \\
&\qquad \qquad  \times \left(\left(\frac{z_{mn}z_{nq}}{z_{mq}}B_n\right)^{-1} \left[\frac{1}{n}\left(\frac{(q^2-m^2)a_n}{n^2-1}+a_m-a_q\right)\right]^n\right)^{\frac{1}{n}}.
\end{align}
As before, the powers of $z_{ij}$ simply correct the three point function, turning it into a correlator of primaries.  Furthermore, it should be noted that the ``correction'' coefficient
\begin{equation}
\left(\left(\frac{z_{mn}z_{nq}}{z_{mq}}B_n\right)^{-1} \left[\frac{1}{n}\left(\frac{(q^2-m^2)a_n}{n^2-1}+a_m-a_q\right)\right]^n\right)^{\frac{1}{n}}
\end{equation}
is $m\leftrightarrow q$ invariant (keeping track of signs in $B_n$ that change).  As an additional consistency check for the expression above, recall that $\ell_{-1/n}$ is 0 when operating on a bare twist.  The bare twist at location $z_n$ corresponds to setting $a_n=0$ for the expression.  Furthermore, the computation lifts to a 2-point function on the cover, necessitating $a_m=a_q$, which therefore sets the three point function to 0, as needed.

Taking the $n^{\rm th}$ root, as always, introduces a phase unless we consider spinless excitations, balancing left and right moving weights such that phases cancel, or operating with several fractional modes to build whole integer total weight added to one side, which we assume henceforth.

We would like to check the above answer and see that it agrees with our previous results \cite{Burrington:2022dii}.  First, we isolate the correction coefficient as
\begin{align}
\left(\left(\frac{z_{mn}z_{nq}}{z_{mq}}B_n\right)^{-1} \left[\frac{1}{n}\left(\frac{(q^2-m^2)a_n}{n^2-1}+a_m-a_q\right)\right]^n\right)^{\frac{1}{n}}.
\end{align}
We may compare to the examples in \cite{Burrington:2022dii}, one may simply take $m=5, n=2, q=4$, and find agreement.

We may now use the $m,n,q$ symmetry to write down all three answers for single, lowest level, fractional excitations on each operator
\begin{align}
& \langle \left[\ell_{-1/m}\mathcal{O}_{m,{-a_m/m}}\sigma_{m}\right](z_m) \left[\mathcal{O}_{n,{-a_n/n}}\sigma_{n}\right](z_n) \left[\mathcal{O}_{q,{-a_{q}/(q)}}\sigma_{q}\right](z_{q})\rangle \nn \\
&=\langle \left[\mathcal{O}_{m,{-a_m/m}}\sigma_{m}\right](z_m) \left[\mathcal{O}_{n,{-a_n/n}}\sigma_{n}\right](z_n) \left[\mathcal{O}_{q,{-a_{q}/(q)}}\sigma_{q}\right](z_{q})\rangle \nn \\
&\qquad \qquad  \times \left(\left(\frac{z_{mn}z_{mq}}{z_{nq}}B_m\right)^{-1} \left[\frac{1}{m}\left(\frac{(n^2-q^2)a_m}{m^2-1}+a_q-a_n\right)\right]^m\right)^{\frac{1}{m}} \\
& \langle \left[\mathcal{O}_{m,{-a_m/m}}\sigma_{m}\right](z_m) \left[\ell_{-1/n}\mathcal{O}_{n,{-a_n/n}}\sigma_{n}\right](z_n) \left[\mathcal{O}_{q,{-a_{q}/(q)}}\sigma_{q}\right](z_{q})\rangle \nn \\
&=\langle \left[\mathcal{O}_{m,{-a_m/m}}\sigma_{m}\right](z_m) \left[\mathcal{O}_{n,{-a_n/n}}\sigma_{n}\right](z_n) \left[\mathcal{O}_{q,{-a_{q}/(q)}}\sigma_{q}\right](z_{q})\rangle \nn \\
&\qquad \qquad  \times \left(\left(\frac{z_{mn}z_{nq}}{z_{mq}}B_n\right)^{-1} \left[\frac{1}{n}\left(\frac{(q^2-m^2)a_n}{n^2-1}+a_m-a_q\right)\right]^n\right)^{\frac{1}{n}}\\
& \langle \left[\mathcal{O}_{m,{-a_m/m}}\sigma_{m}\right](z_m) \left[\mathcal{O}_{n,{-a_n/n}}\sigma_{n}\right](z_n) \left[\ell_{-1/q}\mathcal{O}_{q,{-a_{q}/(q)}}\sigma_{q}\right](z_{q})\rangle \nn \\
&=\langle \left[\mathcal{O}_{m,{-a_m/m}}\sigma_{m}\right](z_m) \left[\mathcal{O}_{n,{-a_n/n}}\sigma_{n}\right](z_n) \left[\mathcal{O}_{q,{-a_{q}/(q)}}\sigma_{q}\right](z_{q})\rangle \nn \\
&\qquad \qquad  \times \left(\left(\frac{z_{mq}z_{nq}}{z_{mn}}B_q\right)^{-1} \left[\frac{1}{q}\left(\frac{(m^2-n^2)a_q}{q^2-1}+a_n-a_m\right)\right]^q\right)^{\frac{1}{q}}.
\end{align}
The most efficient way to generate the other terms is by a cyclic $m\rightarrow n\rightarrow q\rightarrow m$ interchange.

We now move on to a situation where a contour pull will be necessary, and consider the excitation
\begin{align}
&\langle \left[\mathcal{O}_{m,{-a_m/m}}\sigma_{m}\right](z_m) \left[\ell_{-2/n}\mathcal{O}_{n,{-a_n/n}}\sigma_{n}\right](z_n) \left[\mathcal{O}_{q,{-a_{q}/(q)}}\sigma_{q}\right](z_{q})\rangle \nn \\
& \rightarrow m^{-a_m+1} A_{m}^{-a_m/m}n^{-a_n+1} A_{n}^{-a_n/n}q^{-a_q+1} A_{q}^{-a_q/q} \nn \\
& \qquad \times
\langle\mathcal{O}_{m}(t_m) \left[\oint_{t_n}\frac{dt}{2\pi i} (z(t)-z_{n})^{-\frac{2}{n}+1}(\pa z)^{-1}\left(T(t)-\frac{c}{12}\left\{z(t),t\right\}\right)\mathcal{O}_{n}(t_n) \right] \mathcal{O}_{q}(t_q)  \rangle. \label{liftLm2ovn}
\end{align}
The measure of the integral has the expansion
\begin{align}
(z(t)-z_{n})^{-\frac{2}{n}+1}(\pa z)^{-1}&=\left(\frac{z_{mn}z_{nq}}{z_{mq}} B_n\left(\frac{t_{mq}}{t_{mn}t_{nq}}\right)^n\right)^{\frac{-2}{n}} \frac{1}{n} \nn \\
& \qquad \qquad \times (t-t_n)^{-1}\left(1+S_1 (t-t_n)+S_2 (t-t_n)^2+\cdots\right)
\end{align}
with
\begin{align}
S_1&=-\frac{3}{2}\frac{t_{mq}(m^2-q^2)-(n^2-1)(t_{mn}-t_{nq})}{(t_{nq}t_{mn})(n^2-1)} \\
S_2&= \frac{1}{4(n^2-4)(n^2-1)^2t_{mn}^2t_{nq}^2} \nn \\
&\qquad \times \bigg((n^2-4)(n^2-1)^2(3(t_{mn}-t_{nq})^2-(t_{mq})^2)-t_{mq}^2(n^2-1)^2(2(m^2+q^2)-1) \nn \\
&\qquad -6(n^2-1)(n^2-4)(m^2-q^2)t_{mq}(t_{mn}-t_{nq})+3t_{mq}^2(2n^2-5)(m^2-q^2)^2\bigg)\nn \\
\end{align}
Both of these can be seen to be $m\leftrightarrow q$ symmetric, as they must be.  Evaluating the Schwarzian term, we find
\begin{align}
&\oint_{t_n}\frac{dt}{2\pi i}\left(\frac{z_{mn}z_{nq}}{z_{mq}} B_n\left(\frac{t_{mq}}{t_{mn}t_{nq}}\right)^n\right)^{\frac{-2}{n}} \frac{1}{n} \nn \\
& \qquad \qquad \times (t-t_n)^{-1}\left(1+S_1 (t-t_n)+S_2 (t-t_n)^2+\cdots\right)\frac{-c}{12}\left\{z(t),t\right\} \nn \\
&= \left(\frac{z_{mn}z_{nq}}{z_{mq}} B_n\left(\frac{t_{mq}}{t_{mn}t_{nq}}\right)^n\right)^{\frac{-2}{n}} \frac{1}{n}\frac{c}{32}\frac{t_{mq}^2}{t_{nq}^2t_{mn}^2}\frac{3(m^2-q^2)^2-2(n^2-1)(m^2+q^2+n^2-5)}{(n^2-1)(n^2-4)}
\end{align}
and we see all of the $t_{ij}$ dependence will cancel, and so this part by itself makes sense on the base space.  This provides an interesting check, as it is the only part of the computation that is proportional to the central charge, and so must make sense in the base space by itself if we are to get universal relations.

The remaining computation involving $T(t)$ becomes
\begin{align}
& \langle\mathcal{O}_{m}(t_m) \left[\oint_{t_n}\frac{dt}{2\pi i} (z(t)-z_{n})^{-\frac{2}{n}+1}(\pa z)^{-1}T(t)\mathcal{O}_{n}(t_n) \right] \mathcal{O}_{q}(t_q)  \rangle. \nn \\
& = \left(\frac{z_{mn}z_{nq}}{z_{mq}} B_n\left(\frac{t_{mq}}{t_{mn}t_{nq}}\right)^n\right)^{\frac{-2}{n}} \frac{1}{n} \times \nn \\
& \bigg( \langle\mathcal{O}_{m}(t_m) \left[\oint_{t_n}\frac{dt}{2\pi i} (t-t_n)^{-1}T(t)\mathcal{O}_{n}(t_n) \right] \mathcal{O}_{q}(t_q) \rangle \nn \\
& \qquad + \langle\mathcal{O}_{m}(t_m) \left[(S_1{\mc L}_{-1} + S_2{\mc L}_0)\mathcal{O}_{n}\right](t_n)\mathcal{O}_{q}(t_q)\rangle \bigg) \label{Tremain}
\end{align}
In the second term, we may make the replacements (\ref{subsinL0Lm1atn}), and we find
\begin{equation}
(S_1{\mc L}_{-1} + S_2{\mc L}_0)=S_1\left(\frac{a_m+a_n-a_q}{t_{mn}}-\frac{a_n+a_q-a_m}{t_{nq}}\right)+S_2 a_2.
\end{equation}
We evaluate the remaining term with a contour pull,
\begin{align}
& \langle\mathcal{O}_{m}(t_m) \left[\oint_{t_n}\frac{dt}{2\pi i} (t-t_n)^{-1}T(t)\mathcal{O}_{n}(t_n) \right] \mathcal{O}_{q}(t_q)\rangle \nn \\
&=-\langle\left[\left(\frac{1}{t_{mn}}{\mc L}_{-1}-\frac{1}{t_{mn}^2}{\mc L}_0\right)\mathcal{O}_{m}\right](t_m) \mathcal{O}_{n}(t_n)\mathcal{O}_{q}(t_q)\rangle \label{Tmove}\\
& -\langle\mathcal{O}_{m}(t_m) \mathcal{O}_{n}(t_n)\left[\left(-\frac{1}{t_{nq}}{\mc L}_{-1}-\frac{1}{t_{nq}^2}{\mc L}_0\right)\mathcal{O}_{q}\right](t_q)\rangle. \nn
\end{align}
Again, we make the analogous replacements at each of the points, finding the total contribution
\begin{align}\
&\langle\mathcal{O}_{m}(t_m) \left[\oint_{t_n}\frac{dt}{2\pi i} (z(t)-z_{n})^{-\frac{2}{n}+1}(\pa z)^{-1}T(t)\mathcal{O}_{n}(t_n) \right] \mathcal{O}_{q}(t_q)  \rangle \nn \\
&= \langle\mathcal{O}_{m}(t_m) \mathcal{O}_{n}(t_n) \mathcal{O}_{q}(t_q)  \rangle  \left(\frac{z_{mn}z_{nq}}{z_{mq}} B_n\left(\frac{t_{mq}}{t_{mn}t_{nq}}\right)^n\right)^{\frac{-2}{n}} \frac{1}{n} \nn \\
&\times \Bigg(S_1\left(\frac{a_m+a_n-a_q}{t_{mn}}-\frac{a_n+a_q-a_m}{t_{nq}}\right)+S_2 a_n \nn \\
& -\frac{1}{t_{mn}}\left(-\frac{a_m+a_n-a_q}{t_{mn}}-\frac{a_m+a_q-a_n}{t_{mq}}\right)+\frac{1}{t_{mn}^2}a_m \nn \\
& +\frac{1}{t_{nq}}\left(\frac{a_n+a_q-a_m}{t_{nq}}+\frac{a_m+a_q-a_n}{t_{mq}}\right)+\frac{1}{t_{nq}^2}a_q\Bigg) \nn \\
&= \langle\mathcal{O}_{m}(t_m) \mathcal{O}_{n}(t_n) \mathcal{O}_{q}(t_q)  \rangle  \left(\frac{z_{mn}z_{nq}}{z_{mq}} B_n\left(\frac{t_{mq}}{t_{mn}t_{nq}}\right)^n\right)^{\frac{-2}{n}} \frac{1}{n} \frac{t_{mq}^2}{t_{mn}^2t_{nq}^2}  \nn \\
& \times \Bigg(\frac{1}{2}\frac{(3m^2-3q^2+n^2-1)a_q+(3q^2-3m^2+n^2-1)a_m}{(n^2-1)}  \\ &\qquad \qquad +\frac{1}{4}\frac{3(m^2-q^2)^2(2n^2-5)-(n^2-1)^2(2(m^2+q^2)-1)}{(n^2-1)^2(n^2-4)}a_n\Bigg) \nn
\end{align}
We note again that all $t_{ij}$ contributions cancel.  Interestingly, one can use the expression (\ref{Tremain}) added to $(\ref{Tmove})$ and use the global ward identities on the cover
\begin{align}
&\langle [({\mc L}_{-1})\mathcal{O}_{m}](t_m)\mathcal{O}_{n}(t_n)\mathcal{O}_{q}(t_q)\rangle+ {\rm cyclic}=0 \nn \\
&\langle [(t_m{\mc L}_{-1}+{\mc L}_0)\mathcal{O}_{m}](t_m)\mathcal{O}_{n}(t_n)\mathcal{O}_{q}(t_q)\rangle+ {\rm cyclic}=0 \\
&\langle [(t_m^2{\mc L}_{-1}+2t_m{\mc L}_0)\mathcal{O}_{m}](t_m)\mathcal{O}_{n}(t_n)\mathcal{O}_{q}(t_q)\rangle+ {\rm cyclic}=0 \nn
\end{align}
to rewrite the entire correlator in terms of numeric coefficients multiplying ${\mc L}_0$ operators acting at each point, giving the same expression as above.

Putting this all together, we find that the correlator on the base space for the excited operator is directly related to the correlator on the base space for the unexcited ancestor via
\begin{align}
&\langle \left[\mathcal{O}_{m,{-a_m/m}}\sigma_{m}\right](z_m) \left[\ell_{-2/n}\mathcal{O}_{n,{-a_n/n}}\sigma_{n}\right](z_n) \left[\mathcal{O}_{q,{-a_{q}/(q)}}\sigma_{q}\right](z_{q})\rangle \nn \\
&= \langle \left[\mathcal{O}_{m,{-a_m/m}}\sigma_{m}\right](z_m) \left[\mathcal{O}_{n,{-a_n/n}}\sigma_{n}\right](z_n) \left[\mathcal{O}_{q,{-a_{q}/(q)}}\sigma_{q}\right](z_{q})\rangle \nn \\
&\times \left(\frac{z_{mn}z_{nq}}{z_{mq}} B_n\right)^{\frac{-2}{n}} \frac{1}{n} \nn \\
&\times \Bigg(\frac{1}{2}\frac{(3m^2-3q^2+n^2-1)a_q+(3q^2-3m^2+n^2-1)a_m}{(n^2-1)}  \\
&\qquad \qquad +\frac{1}{4}\frac{3(m^2-q^2)^2(2n^2-5)-(n^2-1)^2(2(m^2+q^2)-1)}{(n^2-1)^2(n^2-4)}a_n \nn \\
& \qquad \qquad +\frac{c}{32}\frac{3(m^2-q^2)^2-2(n^2-1)(m^2+q^2+n^2-5)}{(n^2-1)(n^2-4)}\Bigg).
\end{align}
The other two excitations $\ell_{-2/q}$ and $\ell_{-2/m}$ operating at the other points may be obtained by cyclic rearrangement $m\rightarrow n\rightarrow q \rightarrow m$. As before, the powers of $z_{ij}$ appearing in the dressing coefficient simply shift the form of the correlator of primaries with appropriate weights:
\begin{align}
h_{\rm tot,m}&=\frac{c}{24}(m-1/m)+\frac{a_m}{m} \nn \\
h_{\rm tot,n}&=\frac{c}{24}(n-1/n)+\frac{a_n}{n}+\frac{2}{n} \\
h_{\rm tot,q}&=\frac{c}{24}(q-1/q)+\frac{a_q}{q} \nn
\end{align}
We may check the results of our previous work with $m=5, n=2, q= 4$, finding exact agreement with \cite{Burrington:2022dii}.

\section{4-point function $(n)-(2)-(2)-(n)$}
\label{4ptsection}

In the previous section, the $sl(2)$ transformations on the base and the $sl(2)$ transformations on the cover give enough freedom to fix all points of interest: the location of the operators on the base, and the covering space location of the ramified points in the map.  However, more complicated 3-point functions with multi-cycle operators lead to a less clean separation.  However, these 3-point functions can be seen as confluences of higher point functions, and so we consider these captured by such investigations.  Thus, the next most complicated case is considering 4-point functions of single cycle operators, for which we conisder the $(n)-(2)-(2)-(n)$ as an example.

We consider first the map
\begin{equation}
w=-u^n\frac{((n-1)s-n)u-(ns-(n+1))}{s u-1}
\end{equation}
where $s$ is a parameter of the map (this map was first considered in \cite{Lunin:2000yv} in a slightly different form \footnote{To relate the variables in \cite{Lunin:2000yv} to ours, set their map variable to $t_{\rm LM}$, and $u=t_{\rm LM}/s, a_{\rm LM}=(s(ns-n-1)/(ns-n-s))$, where $t_{LM}$ is the coordinate and $a_{\rm LM}$ is the map data used in \cite{Lunin:2000yv}.  The branch cuts are taken such that the variables in \cite{Lunin:2000yv} are given by $t_+=s$, $t_-=(ns-n-1)/(ns-n-s)$}).  Taking the derivative, we find
\begin{equation}
\frac{dw}{du}=-nu^{n-1}\frac{(u-1)((n-1)s^2 u-n(u+1)s+(n+1)}{(su-1)^2}
\end{equation}
The ramified points are $u=0$ and $u=\infty$ each with ramification $n-1$, and $u=1$ with ramification 1, and
\begin{equation}
u=u_s=\frac{ns-(n+1)}{s((n-1)s-n)} \label{usdef}
\end{equation}
with ramification 1.  The point $u=1/s$ also looks interesting, however this is just the unramified image of $w=\infty$ on the $n+1$ sheeted cover.  The parameter $s$ simply controls the value of the location of the $4^{\rm th}$ ramified point in the map, the location of the second 2-cycle twist operator.  This can be thought of as the cross ratio
\begin{equation}
\zeta_u=\lim_{u_0\rightarrow 0, u_1\rightarrow 1, u_\infty\rightarrow \infty}
\frac{(u_1-u_\infty)(u_s-u_0)}{(u_s-u_\infty)(u_1-u_0)}=u_s
\end{equation}
In addition, one can see the images $u=0 \rightarrow w=0$, $u=1 \rightarrow w=1$, $u=\infty \rightarrow w=\infty$ and $u=u_s \rightarrow w=w_s$ with
\begin{equation}
w_s=\frac{1}{s^{n+1}} \frac{(ns-(n+1))^{n+1})}{((n-1)s-n)^{n-1}} \label{wsdef}
\end{equation}
which can be thought of the cross ratio in the $w$ base space
\begin{equation}
\zeta_w=\lim_{w_0\rightarrow 0, w_1\rightarrow 1, w_\infty\rightarrow \infty}
\frac{(w_1-w_\infty)(w_s-w_0)}{(w_s-w_\infty)(w_1-w_0)}=w_s
\end{equation}
Thus, the single parameter $s$ controls both the cross ratio of ramified points on the cover, and the cross ratio of the location of operators on the base.  One can take the relation (\ref{usdef}) and invert to find $s$, giving two solutions.  However, this has very little meaning on the base.  We prefer to parameterized everything in terms of $s$, which specifies the map without branch cut ambiguities, and defines the cross ratio on both the $u$-plane and the $w$-plane.

As before, we wish to map all points to finite points, and do so with the maps
\begin{equation}
u=\frac{(t_{1}-t_{\infty})}{(t_1-t_0)}\frac{(t-t_0)}{(t-t_\infty)}
\end{equation}
where $t_i$ are the ramified points in the new map $w(u(t))$ (the subscripts on $t_i$ give the original locations in the $u$-plane).  The image of $u_s$ in the $t$ plane given by the equation
\begin{equation}
u_s=\frac{(t_{1}-t_{\infty})}{(t_1-t_0)}\frac{(t_s-t_0)}{(t_s-t_\infty)}=\zeta_t=\frac{ns-(n+1)}{s((n-1)s-n)} \label{zetat}
\end{equation}
and so this is still interpreted as the cross ratio.  Explicitly solving for $t_s$, one finds
\begin{equation}
t_s=\frac{t_0(t_1-t_\infty)(n-1)s^2-n(t_1 t_0 + t_1 t_\infty - 2 t_0t_\infty)s+t_\infty(t_1-t_0)(n+1)}{(t_1-t_\infty)(n-1)s^2-n(2t_1-t_0-t_\infty)s+(t_1-t_0)(n+1)}
\end{equation}

We likewise map to the $z$ plane via
\begin{equation}
w=\frac{(z_1-z_\infty)}{(z_1-z_0)}\frac{(z-z_0)}{(z-z_\infty)}
\end{equation}
where the subscripts on the $z_i$ denote the original locations of the operators in the $w$ plane.  The location of the 4-th operator is given by $z_s$ which is found by
\begin{equation}
w_s=\frac{(z_1-z_\infty)}{(z_1-z_0)}\frac{(z_s-z_0)}{(z_s-z_\infty)}=\zeta_z= \frac{1}{s^{n+1}} \frac{(ns-(n+1))^{n+1})}{((n-1)s-n)^{n-1}} \label{zetaz}
\end{equation}
so, again, this cross ratio is controlled entirely in terms of $s$. Above, one can imagine inverting (\ref{zetaz}) to find $s$ in terms of $\zeta_z$.  In some sense this is true, however, there will be $2n$ solutions to the above equation related by choice of branch cuts.  Solving for $z$ we find
\begin{equation}
z=\frac{z_\infty(z_0-z_1)w+z_0(z_1-z_\infty)} {(z_0-z_1)w+(z_1-z_\infty)}
\end{equation}
and
\begin{equation}
z_s=\frac{z_\infty(z_0-z_1)(ns-(n+1))^{n+1})+z_0(z_1-z_\infty)s^{n+1}((n-1)s-n)^{n-1}} {(z_0-z_\infty)(ns-(n+1))^{n+1})+(z_1-z_\infty)s^{n+1}((n-1)s-n)^{n-1}}
\end{equation}

Expanding near various points, we find
\begin{align}
& (z-z_0)=\frac{(z_0-z_\infty)(z_0-z_1)}{(z_\infty-z_1)}w \frac{1}{\left(1-\frac{(z_0-z_1)}{(z_\infty-z_1)}w\right)}=(z_0-z_\infty)\sum_{i=1}^\infty \left(\frac{(z_0-z_1)}{(z_\infty-z_1)}w\right)^i \nn \\
& (z-z_\infty)=\frac{(z_\infty-z_0)(z_1-z_\infty)}{(z_1-z_0)}\frac{1}{w} \frac{1}{\left(1-\frac{(z_1-z_\infty)}{(z_1-z_0)}\frac{1}{w}\right)}=(z_\infty-z_0)\sum_{i=1}^\infty \left(\frac{(z_1-z_\infty)}{(z_1-z_0)}\frac{1}{w}\right)^i  \\
& (z-z_1) = \frac{(z_1-z_\infty)(z_1-z_0)}{(z_0-z_\infty)} (w-1) \frac{1}{\left(1-\frac{(z_1-z_0)}{(z_0-z_\infty)}(w-1)\right)} \nn \\
& \qquad \qquad \qquad \qquad \qquad \qquad \qquad \qquad \qquad \qquad = (z_1-z_\infty)\sum_{i=1}^\infty \left(\frac{(z_1-z_0)}{(z_0-z_\infty)}(w-1)\right)^i \nn \\
& (z-z_s) =\frac{(z_s-z_\infty)^2(z_1-z_0)}{(z_1-z_\infty)(z_0-z_\infty)}(w-w_s)\frac{1} {\left(1-\frac{(z_s-z_\infty)(z_1-z_0)}{(z_1-z_\infty)(z_0-z_\infty)}(w-w_s)\right)} \nn \\
&\qquad \qquad \qquad \qquad \qquad \qquad \qquad = (z_s-z_\infty) \sum_{i=1}^\infty \left(\frac{(z_s-z_\infty)(z_1-z_0)}{(z_1-z_\infty)(z_0-z_\infty)}(w-w_s)\right)^i \nn.
\end{align}
Using the above expressions helps organize the calculations.  We first expand in $w$ near the relevant values, keeping in mind the subsequent expansions in $u$ and $t$.  Since the map $w(u)-w(u_i)\sim (u-u_i)^{n_i}$ is ramified, higher powers in $w$ in the above expansion only become important when expanding sufficiently far in $(u-u_i)$, namely every $n_i$ powers.  Since $u-u_i$ approaches the appropriate value linearly in $t-t_i$, this means that the expansion may be truncated appropriately for finding the $t$ expansion to this level as well.  In short, if one wants to obtain the first $k$ terms in the $t$ expansion of $z(t)$, one must keep up to the $1+\lfloor{k/n_i}\rfloor$ term in $w$ expansion of $z(w)$.  Additionally, the higher powers in $w$ also come with new factors of $z_{ij}$, which should be kept separate for comparison on the base space.

We now turn to the question of how the four point function on the cover gives rise to a 4-point function on the base.  As usual, to simplify discusion, we restrict to the holomorphic part of the correlator, and consider the four-point function on the base
\begin{equation}
\langle \mathcal{O}_{n,{-a_\infty/n}}\sigma_{n}(z_\infty)\mathcal{O}_{1,{-a_1/2}}\sigma_{2}(z_1) \mathcal{O}_{s,{-a_s/2}}\sigma_{2}(z_s)\mathcal{O}_{0,{-a_0/n}}\sigma_{n}(z_0)\rangle
\end{equation}
which lifts to a four-point function of primaries on the cover.  We expand near the following points with
\begin{align}
&(z-z_0)=A_0 (t-t_0)^n \bigg(1+S_{1,0}(t-t_0) + S_{2,0}(t-t_0)^2+\cdots\bigg) \nn \\
& (z-z_\infty)= A_\infty (t-t_\infty)^n \bigg(1+S_{1,\infty}(t-t_\infty) + S_{2,\infty}(t-t_\infty)^2+\cdots\bigg) \nn \\
& (z-z_1)=A_1(t-t_1)^2 \bigg(1+S_{1,1}(t-t_1) + S_{2,1}(t-t_1)^2+\cdots\bigg) \nn \\
& (z-z_s)=A_s(t-t_s)^2 \bigg(1+S_{1,s}(t-t_s)+S_{2,s}(t-t_s)^2+\cdots\bigg)
\end{align}
where
\begin{align}
&A_0=\frac{(z_0-z_\infty)(z_0-z_1)}{(z_\infty-z_1)}\left(\frac{(t_\infty-t_1)}{(t_0-t_\infty)(t_0-t_1)}\right)^n(-1)(ns-(n+1)) \nn \\
&A_\infty=\frac{(z_\infty-z_0)(z_1-z_\infty)}{(z_1-z_0)} \left(\frac{(t_1-t_0)}{(t_\infty-t_0)(t_1-t_\infty)}\right)^n\frac{(-1)s}{(n-1)s-n}  \\
&A_1=\frac{(z_1-z_\infty)(z_1-z_0)}{(z_0-z_\infty)} \left(\frac{(t_0-t_\infty)}{(t_1-t_\infty)(t_1-t_0)}\right)^2 \frac{(-1)n((n-1)s-(n+1))}{2(s-1)} \nn \\
&A_s=\frac{(z_s-z_\infty)^2(z_1-z_0)}{(z_1-z_\infty)(z_0-z_\infty)}  \nn \\
& \quad \times \left(\frac{(t_1-t_{\infty})(t_0-t_\infty)}{(t_s-t_\infty)^2(t_1-t_0)}\right)^2 \frac{ns((n-1)s-(n+1))(ns-(n+1))^{n-1}}{2(s-1)s^n((n-1)s-n)^{n-3}}. \\
\end{align}
The further terms in the expansions are parameterized by $S_{{\rm or},i}$ (subscripts given as ``order of expansion, location''), which appear in the appendix \ref{expansionAppx}.  However, the leading order terms of the above expansions are highly suggestive of how to organize the calculation moving forward.

First, we note that the points $z_0, z_1, z_\infty$ have been treated differently, in that they are directly controlled by the $sl(2)$ transformation on the base which leads to the above powers in the expansion.  The fourth point, while important, has been controlled by the parameter $s$, which appears implicitly in $z_s, w_s=\zeta_z, u_s=\zeta_t, t_s$.  Thus, we treat these points differently in the expression for the 4-point functions, both on the base space and on the cover.  Considering $4$ primaries with weights $h_i$, we may write the four point functions as
\begin{align}
&\langle \mathcal{O}_{\infty}(z_\infty)\mathcal{O}_{1}(z_1) \mathcal{O}_{s}(z_s)\mathcal{O}_{0}(z_0)\rangle \nn \\
&\quad = \left(\frac{z_{\infty,0}z_{1,\infty}}{z_{1,0}}\right)^{-h_\infty} \left(\frac{z_{0,\infty}z_{0,1}}{z_{\infty,1}} \right)^{-h_0}
\left(\frac{z_{1,\infty}z_{1,0}}{z_{0,\infty}}\right)^{-h_1}
\left(\frac{(z_{\infty,s})^2z_{0,1}}{z_{1,\infty}z_{0,\infty}}\right)^{-h_s} f_z(\zeta_z).
\end{align}
We use this convention to define the function of the cross ratio on the base space $f_z(\zeta_z)$.  We apply a similar convention in the covering space, so that
\begin{align}
&\langle \mathcal{O}_{\infty}(t_\infty)\mathcal{O}_{1}(t_1) \mathcal{O}_{s}(t_s)\mathcal{O}_{0}(t_0)\rangle \nn \\
&\quad = \left(\frac{t_{\infty,0}t_{1,\infty}}{t_{1,0}}\right)^{-a_\infty} \left(\frac{t_{0,\infty}t_{0,1}}{t_{\infty,1}} \right)^{-a_0}
\left(\frac{t_{1,\infty}t_{1,0}}{t_{0,\infty}}\right)^{-a_1}
\left(\frac{(t_{\infty,s})^2t_{0,1}}{t_{1,\infty}t_{0,\infty}}\right)^{-a_s} f_t(\zeta_t)
\end{align}
is used to define $f_t(\zeta_t)$ on the cover.  These conventions are different from \cite{DiFrancesco:1997nk}, for example, but only differ by absorbing some function of the cross ratio into the definitions of $f_t$ and $f_z$.

Using this convention, we find that in the lifted computation on the cover all of the explicit powers of $t_{i,j}$ cancel, leaving
\begin{align}
&\langle \mathcal{O}_{n,{-a_\infty/n}}\sigma_{2}(z_\infty)\mathcal{O}_{1,{-a_1/2}}\sigma_{2}(z_1) \mathcal{O}_{s,{-a_s/2}}\sigma_{2}(z_s)\mathcal{O}_{0,{-a_0/n}}\sigma_{n}(z_0)\rangle \nn \\
&\rightarrow  \left(\frac{z_{\infty,0}z_{1,\infty}}{z_{1,0}}\right)^{-a_\infty/n} \left(\frac{z_{0,\infty}z_{0,1}}{z_{\infty,1}} \right)^{-a_0/n}
\left(\frac{z_{1,\infty}z_{1,0}}{z_{0,\infty}}\right)^{-a_1/2}
\left(\frac{(z_{\infty,s})^2z_{0,1}}{z_{1,\infty}z_{0,\infty}}\right)^{-a_s/2} \nn \\
& \times  \frac{\left((-1)(ns-(n+1))\right)^{-a_0/n}}{n} \frac{\left(\frac{(-1)s}{(n-1)s-n}\right)^{-a_\infty/n}}{n} \nn \\
& \times \frac{\left(\frac{(-1)n((n-1)s-(n+1))}{2(s-1)}\right)^{-a_1/2}}{2} \frac{\left(\frac{ns((n-1)s-(n+1))(ns-(n+1))^{n-1}}{2(s-1)s^n((n-1)s-n)^{n-3}}\right)^{-a_s/2}}{2} f_t(\zeta_t). \label{simple4ptexcite}
\end{align}
We see above that the powers of $z_{ij}$ are the correct factors to shift the powers of $z_{ij}$ to the correct ones for a 4-point function of primaries on the base space, i.e.
\begin{equation}
h_{tot,i}=\frac{c}{24}(n_i-1/n_i)+ a_i/n_i.
\end{equation}
Furthermore, we have that the cross ratio on the base space is given by
\begin{equation}
\zeta_z= \frac{1}{s^{n+1}} \frac{(ns-(n+1))^{n+1})}{((n-1)s-n)^{n-1}} \label{zetazdef}
\end{equation}
and that the cross ratio on the cover is given by
\begin{equation}
\zeta_t=\frac{ns-(n+1)}{s((n-1)s-n)}
\end{equation}
both of which are solely functions of $s$ (and the specified $n$).  So, the remaining question is whether the above calculation makes sense in the base space, a problem that exists even for the four point function of {\it bare} twists in \cite{Lunin:2000yv}.  We will argue that it does, simply because only $s$ appears.

First, one may ask the question for how many values of $s$ does one get the same value of $\zeta_z$?  It is clear the relation (\ref{zetazdef}) will give an order $2n$ polynomial in $s$ given $\zeta_z$.  Thus, there are generically $2n$ solutions that all give the same $\zeta_z$.  How are we to determine which of these is relevant?  In previous work \cite{Lunin:2000yv,Burrington:2018upk,DeBeer:2019oxm} it has been shown that different limits of covering map parameters in fact leads to different exchange channels being explored (for this exact same map, discussed somewhat differently).  Some of these are realized by the same limit on the base space, e.g. there are two limits for $s$ where $z_s\rightarrow z_0$ and these contain the two fusions $(2)-(n) \rightarrow (n-1)$ and $(2)-(n)\rightarrow (n+1)$ as exchange channels \cite{Burrington:2018upk}.  Let us consider first the $(2)-(n) \rightarrow (n+1)$ fusion.  This corresponds to a group product with a representative $(1,2,3,\cdots ,n)(n,n+1)=(1,2,3,4,\cdots,n,n+1)$.  However, when transporting the twist $n$ operator around the twist $2$ operator, back to the same point, the branch cuts that define the simply connected patch on the base space interfere with each other.  Such a transport conjugates each of the cycles $(1,2,3,\cdots ,n)$ and $(n,n+1)$ by $(1,2,3,4,\cdots,n,n+1)$, leaving the total group product unchanged, but using different conjugacy class representatives in the calculation.  Repeating this, one gets a set of $n+1$ conjugacy class equivalent calculations.  This is seen on the cover via roots of $w_s^{1/(n+1)}$ appearing in the OPE limit \cite{Burrington:2018upk,DeBeer:2019oxm}.  A similar argument for the $(2)-(n) \rightarrow (n-1)$ fusion applies, implying a total of $n+1+n-1=2n$ total exchange channels.  Thus, when solving (\ref{zetazdef}) for $s$, we interpret each of the $2n$ solutions as one of the $2n$ possible exchange channels when taking OPE limits of the 4-point function.  Thus, $s$ not only specifies the cross ratio, but which of the exchange channels is being considered, offering more information.  This allows for checks of specific crossing channels individually, as in \cite{Lunin:2000yv,Burrington:2018upk,DeBeer:2019oxm}.  However, this information is inherently available from the base space point of view: different exchange channels necessarily arise in different conformal blocks, and these different functional forms are detectable in OPE limits.  One may be concerned that not all crossing channels appear, and there may be a ``disconnected sum'', however, some exploration of which conjugacy class representatives are accessible using parallel transport has been undertaken \cite{Dei:2019iym}, and this appears to connect all conjugacy class representatives into one computation related by analytic continuation (corresponding to the $2n$ solutions mentioned here).  In the end, one presumably sums over all such images, making the final answer not depend on which branch of the solution to (\ref{zetazdef}) is taken (therefore giving a sum of all crossing channels in the OPE limit, as should be expected). The above comments inform the calculations of  \cite{Lunin:2000yv,Burrington:2018upk,DeBeer:2019oxm} when checking the crossing channels.  The fact that there are branch cuts that appear in individual terms we consider to be expected, given that we are working with twist operators which have not yet been summed to make orbifold invariant operators.

Thus, henceforth we take that reducing the problems to functions of $s$ (and $n$) and $f_t(\zeta_t(s))$ is tantamount to having an acceptable function on the base space, the full answer being the sum over all such images of the solution to (\ref{zetazdef}).  Expressions of the type (\ref{simple4ptexcite}) are to be understood for one choice of group elements in the conjugacy class and so the $s$ appearing there is a specific solution to (\ref{zetazdef}), furnishing the function of $s$ that needs to be summed over solutions to (\ref{zetazdef}).

We now wish to consider excitations using fractional virasoro modes, some of which will require the Schwarzian, which may be written
\begin{align}
\{z,t\}=\{w,u\}u'^2 & =-\frac{n^2-1}{2(t-t_\infty)^2}-\frac{n^2-1}{2(t-t_0)^2}-\frac{3}{2(t-t_1)^2}-\frac{3}{2(t-t_s)^2} \nn \\
&\qquad +\frac{F_{\infty,0}}{(t-t_\infty)(t-t_0)}+\frac{F_{\infty,1}}{(t-t_\infty)(t-t_1)}+\frac{F_{\infty,s}}{(t-t_\infty)(t-t_s)} \nn \\
&\qquad +\frac{F_{0,1}}{(t-t_0)(t-t_1)}+\frac{F_{0,s}}{(t-t_0)(t-t_s)} +\frac{F_{1,s}}{(t-t_1)(t-t_s)}
\end{align}
Where $F_{i,j}$ are a set of functions, which we require to be functions only of $s$ and $n$.  This constraint gives the equations
\begin{align}
& F_{\infty,0}=n^2-4+F_{1,s}, \qquad F_{\infty,1}=F_{0,s}, \qquad F_{\infty,w}=F_{0,1}=-F_{0,s}-F_{1,s}+3, \nn \\
& (n+1)(-n-2+F_{0,s}+F_{1,s})s^2+(2n^2+3n-2-2nF_{0,s}-nF_{1,s})s \nn \\
& \qquad \qquad \qquad \qquad \qquad \qquad \qquad \qquad \qquad+(n-1)(-n-1+ F_{0,s})=0
\end{align}
where $F_{0,s}$ and $F_{1,s}$ are used to specify all other functions, and are freely specifiable up to the constraint in the last line above.  We note that the double pole terms only depend on the size of the twists associated with the pole terms.  If we enforce a similar condition on the other terms, this would require that $F_{\infty,1}=F_{\infty,s}=F_{0,1}=F_{0,s}$ because each of these terms is associated with a mixed twist-$n$/twist-2 poles.  This constraint may be enforced, which then gives a fixed set of coefficients
\begin{align}
F_{\infty,1}=F_{\infty,s}=F_{0,1}&=F_{0,s}=\frac{(s-1)^2(n^2-1)}{(n-1)s^2-(n+1)} \nn \\
F_{1,s}= -2\frac{(s-1)^2(n^2-1)}{(n-1)s^2-(n+1)}+3, &\qquad F_{\infty,0}=-2\frac{(s-1)^2(n^2-1)}{(n-1)s^2-(n+1)}+n^2-1
\end{align}

We turn to the lowest virasoro excitation modes
\begin{align}
&\langle [\ell_{-1/n}\mathcal{O}_{\infty,{-a_\infty/n}}\sigma_{n}](z_\infty)\mathcal{O}_{1,{-a_1/2}}\sigma_{2}(z_1) \mathcal{O}_{s,{-a_s/2}}\sigma_{2}(z_s)\mathcal{O}_{0,{-a_0/n}}\sigma_{n}(z_0)\rangle, \nn \\
&\langle \mathcal{O}_{\infty,{-a_\infty/n}}\sigma_{n}(z_\infty)[\ell_{-1/2}\mathcal{O}_{1,{-a_1/2}}\sigma_{2}](z_1) \mathcal{O}_{s,{-a_s/2}}\sigma_{2}(z_s)\mathcal{O}_{0,{-a_0/n}}\sigma_{n}(z_0)\rangle, \\
&\langle \mathcal{O}_{\infty,{-a_\infty/n}}\sigma_{n}(z_\infty)\mathcal{O}_{1,{-a_1/2}}\sigma_{2}(z_1) [\ell_{-1/2}\mathcal{O}_{s,{-a_s/2}}\sigma_{2}](z_s)\mathcal{O}_{0,{-a_0/n}}\sigma_{n}(z_0)\rangle,\nn \\
&\langle \mathcal{O}_{\infty,{-a_\infty/n}}\sigma_{n}(z_\infty)\mathcal{O}_{1,{-a_1/2}}\sigma_{2}(z_1) \mathcal{O}_{s,{-a_s/2}}\sigma_{2}(z_s)[\ell_{-1/n}\mathcal{O}_{0,{-a_0/n}}\sigma_{n}(z_0)]\rangle.\nn
\end{align}
Looking at the first of these expressions, we see that the correction
\begin{align}
&\frac{\langle [\ell_{-1/n}\mathcal{O}_{\infty,{-a_\infty/n}}\sigma_{2}](z_\infty)\mathcal{O}_{1,{-a_1/2}}\sigma_{2}(z_1) \mathcal{O}_{s,{-a_s/2}}\sigma_{2}(z_s)\mathcal{O}_{0,{-a_0/n}}\sigma_{n}(z_0)\rangle}{\langle \mathcal{O}_{\infty,{-a_\infty/n}}\sigma_{2}(z_\infty)\mathcal{O}_{1,{-a_1/2}}\sigma_{2}(z_1) \mathcal{O}_{s,{-a_s/2}}\sigma_{2}(z_s)\mathcal{O}_{0,{-a_0/n}}\sigma_{n}(z_0)\rangle}\nn \\
& = \frac{\langle\oint \frac{dt}{2\pi i} (z(t)-z_{\infty})^{\frac{-1}{n}+1}\frac{1}{\pa z(t)} \left(T(t)-\frac{c}{12}\{z,t\}\right)\mathcal{O}_{\infty}(t_\infty) \mathcal{O}_{1}(t_1) \mathcal{O}_{s}(t_s)\mathcal{O}_{0}(t_0)\rangle}{\langle \mathcal{O}_{\infty}(t_\infty) \mathcal{O}_{1}(t_1) \mathcal{O}_{s}(t_s)\mathcal{O}_{0}(t_0)\rangle} \nn \\
&=A_\infty^{-1/n}\frac{1}{n} \frac{\langle \left({\mathcal L}_{-1}-\frac{2}{n} S_{1,\infty}{\mathcal L}_0\right)\mathcal{O}_{\infty}(t_\infty) \mathcal{O}_{1}(t_1) \mathcal{O}_{s}(t_s)\mathcal{O}_{0}(t_0)\rangle}{\langle \mathcal{O}_{\infty}(t_\infty) \mathcal{O}_{1}(t_1) \mathcal{O}_{s}(t_s)\mathcal{O}_{0}(t_0)\rangle}
\end{align}
where above we have used that the Schwarzian term does not contribute (which can be checked, keeping in mind that $t_s$ is a known function of $t_0,t_\infty, t_1$ and $s$). The term ${\mathcal L}_{-1}$ is a derivative acting on a constrained functional form divided by this same functional form, and so we find the replacements
\begin{align}
&{\mathcal L}_{-1}\rightarrow \frac{-a_\infty-a_0+a_1+a_s}{t_{\infty,0}}+\frac{a_0-a_1-a_\infty+a_s}{t_{\infty,1}}+\frac{2 a_s}{t_{s}-t_i}+\frac{f_t'}{f}\frac{(t_0-t_s)(t_1-t_s)}{(t_\infty-t_s)^2(t_0-t_1)} \nn \\
&{\mathcal L}_0 \rightarrow a_{\infty}
\end{align}
giving
\begin{align}
&\frac{\langle [\ell_{-1/n}\mathcal{O}_{\infty,{-a_\infty/n}}\sigma_{2}](z_\infty)\mathcal{O}_{1,{-a_1/2}}\sigma_{2}(z_1) \mathcal{O}_{s,{-a_s/2}}\sigma_{2}(z_s)\mathcal{O}_{0,{-a_0/n}}\sigma_{n}(z_0)\rangle}{\langle \mathcal{O}_{\infty,{-a_\infty/n}}\sigma_{2}(z_\infty)\mathcal{O}_{1,{-a_1/2}}\sigma_{2}(z_1) \mathcal{O}_{s,{-a_s/2}}\sigma_{2}(z_s)\mathcal{O}_{0,{-a_0/n}}\sigma_{n}(z_0)\rangle}\nn \\
&= A_\infty^{-1/n}\frac{1}{n}\frac{t_{\infty,0}t_{1,\infty}}{t_{1,0}} \nn \\
&\times \bigg(-a_0+a_1+a_\infty\frac{ns^2-ns-3s^2+4s-2}{s(ns-n-s)}-a_s\frac{ns^2-3ns-s^2+2n+2}{s(ns-n-s)} \nn \\
& \qquad \qquad \qquad -\frac{f_t'(\zeta_t(s))}{f_t(\zeta_t(s))}\frac{(s-1)(ns-n-1)(ns-n-s-1)}{s^2(ns-n-s)^2}\bigg) \label{4ptRatio1}
\end{align}
Note that the explicit powers of $t_{ij}$ that remain cancel those in $A_\infty^{-1/n}$, leaving behind an answer which only depends on $s$.  The powers of $z_{ij}$ on the right hand side above are just so as to shift the weight of the primary
\begin{equation}
h_{\infty,tot}= \frac{c}{24}(n-1/n)+\frac{a_\infty}{n}+\frac{1}{n}.
\end{equation}

The expression (\ref{4ptRatio1}) is sufficiently complicated that a check is in order.  To do so, we examine the case $a_\infty=0$, in which case the operator at $z_\infty$ is a bare twist, on which $\ell_{-1/n}$ must give 0.  In this case, the lift to the cover is that of a 3-point function, instead of a 4-point function, constraining the function $f_t(\zeta_t)$ such that all occurrences of $t_\infty$ must vanish.  This insists on the form
\begin{align}
f_{t,{\rm 3pt}}&=f_0 \left(\frac{t_{0,\infty}t_{1,s}}{t_{\infty,1}t_{0,s}}\right)^{a_0} \left(\frac{t_{1,\infty}t_{0,s}}{t_{0,\infty}t_{1,s}}\right)^{a_1}\left(\frac{t_{\infty,s}t_{0,1}}{t_{\infty,1}t_{0,s}}\frac{t_{\infty,s}t_{0,1}}{t_{\infty,0}t_{1,s}}\right)^{a_s} \nn \\
&=f_0 \left(\frac{1-\zeta_t}{\zeta_t}\right)^{a_0} \left(\frac{\zeta_t}{\zeta_t-1}\right)^{a_1}\left(\frac{1}{\zeta_t}\frac{1}{1-\zeta_t}\right)^{a_s}
\end{align}
with $f_0$ constant, making
\begin{equation}
\frac{f_t'}{f}=\frac{-2a_s \zeta_t+a_0-a_1+a_s}{\zeta_t(\zeta_t-1)}
\end{equation}
and so
\begin{align}
-\frac{f_t'(\zeta_t(s))}{f_t(\zeta_t(s))}\frac{(s-1)(ns-n-1)(ns-n-s-1)}{s^2(ns-n-s)^2}=a_0-a_1+a_s\frac{ns^2-3ns-s^2+2n+2}{s(ns-n-s)}
\end{align}
and so the 4-point function goes to 0 in this special case, completing the check.

Simplifying the expression (\ref{4ptRatio1}), we find
\begin{align}
&\frac{\langle [\ell_{-1/n}\mathcal{O}_{\infty,{-a_\infty/n}}\sigma_{2}](z_\infty)\mathcal{O}_{1,{-a_1/2}}\sigma_{2}(z_1) \mathcal{O}_{s,{-a_s/2}}\sigma_{2}(z_s)\mathcal{O}_{0,{-a_0/n}}\sigma_{n}(z_0)\rangle}{\langle \mathcal{O}_{\infty,{-a_\infty/n}}\sigma_{2}(z_\infty)\mathcal{O}_{1,{-a_1/2}}\sigma_{2}(z_1) \mathcal{O}_{s,{-a_s/2}}\sigma_{2}(z_s)\mathcal{O}_{0,{-a_0/n}}\sigma_{n}(z_0)\rangle}\nn \\
&= \left(\frac{(z_\infty-z_0)(z_1-z_\infty)}{(z_1-z_0)}\frac{(-1)s}{(n-1)s-n}\right)^{\frac{-1}{n}} \nn \\
&\times \bigg(-a_0+a_1+a_\infty\frac{ns^2-ns-3s^2+4s-2}{s(ns-n-s)}-a_s\frac{ns^2-3ns-s^2+2n+2}{s(ns-n-s)} \nn \\
& \qquad \qquad \qquad -\frac{f_t'(\zeta_t(s))}{f_t(\zeta_t(s))}\frac{(s-1)(ns-n-1)(ns-n-s-1)}{s^2(ns-n-s)^2}\bigg)
\end{align}
Thus, since $s$ parameterizes the cross ratio (and which exchange channel is being considered), the above expression makes sense on the base space.  Alternately, one can solve for $s$ in terms of the cross ratio on the base $\zeta_z$, and express the above answer in terms of the cross ratio $s(\zeta_z)$ along with certain choices for branch cuts.  In such a presentation, one can take expressions of the above kind and replace occurrences of $f_t'(t_s(s))$ with $\zeta_z$ derivatives, noting that the term accompanying $f_t'$ is multiplied only by functions of $s$.  This may make the sum over orbifold images somewhat easier to find on the base space.

Now we turn our attention to a situation where contour pulls become advantageous, and explore
\begin{align}
&\langle [\ell_{-2/n}\mathcal{O}_{\infty,{-a_\infty/n}}\sigma_{2}](z_\infty)\mathcal{O}_{1,{-a_1/2}}\sigma_{2}(z_1) \mathcal{O}_{s,{-a_s/2}}\sigma_{2}(z_s)\mathcal{O}_{0,{-a_0/n}}\sigma_{n}(z_0)\rangle
\end{align}
Lifting the computation, and dividing by the parent, we get the equivalence
\begin{align}
&\frac{\langle [\ell_{-2/n}\mathcal{O}_{\infty,{-a_\infty/n}}\sigma_{2}](z_\infty)\mathcal{O}_{1,{-a_1/2}}\sigma_{2}(z_1) \mathcal{O}_{s,{-a_s/2}}\sigma_{2}(z_s)\mathcal{O}_{0,{-a_0/n}}\sigma_{n}(z_0)\rangle}{\langle \mathcal{O}_{\infty,{-a_\infty/n}}\sigma_{2}(z_\infty)\mathcal{O}_{1,{-a_1/2}}\sigma_{2}(z_1) \mathcal{O}_{s,{-a_s/2}}\sigma_{2}(z_s)\mathcal{O}_{0,{-a_0/n}}\sigma_{n}(z_0)\rangle} \nn \\
& = \frac{\langle\oint \frac{dt}{2\pi i} (z(t)-z_{\infty})^{\frac{-2}{n}+1}\frac{1}{\pa z(t)} \left(T(t)-\frac{c}{12}\{z,t\}\right)\mathcal{O}_{\infty}(t_\infty) \mathcal{O}_{1}(t_1) \mathcal{O}_{s}(t_s)\mathcal{O}_{0}(t_0)\rangle}{\langle \mathcal{O}_{\infty}(t_\infty) \mathcal{O}_{1}(t_1) \mathcal{O}_{s}(t_s)\mathcal{O}_{0}(t_0)\rangle} \nn \\
&=A_\infty^{-2/n}\frac{1}{n} \frac{1}{{\langle \mathcal{O}_{\infty}(t_\infty) \mathcal{O}_{1}(t_1) \mathcal{O}_{s}(t_s)\mathcal{O}_{0}(t_0)\rangle}} \nn \\
&\times \Bigg[\langle\left(\oint \frac{dt}{2\pi i} (t-t_\infty)^{-1}T(t)\right)\mathcal{O}_{\infty}(t_\infty)\mathcal{O}_{1}(t_1) \mathcal{O}_{s}(t_s)\mathcal{O}_{0}(t_0)\rangle \nn \\
& + \langle\left(-\frac{3}{n} S_{1,\infty}{\mathcal L}_{-1} +\frac{(2n+5)S_{1,\infty}^2-4n S_{2,\infty}}{n^2}L_{0}\right)\mathcal{O}_{\infty}(t_\infty)\mathcal{O}_{1}(t_1) \mathcal{O}_{s}(t_s)\mathcal{O}_{0}(t_0)\rangle\Bigg] \nn \\
& -\frac{c}{12}A_\infty^{-2/n}\frac{1}{n} \frac{(t_1-t_0)^2}{(t_1-t_\infty)^2(t_0-t_\infty)^2}\frac{3(s-1)^2((n-1)s^2-(n+1))}{((n-1)s-n)^2s^2}
\end{align}
Pulling the contours, and replacing operators as before, we find
\begin{align}
&\frac{\langle [\ell_{-2/n}\mathcal{O}_{\infty,{-a_\infty/n}}\sigma_{2}](z_\infty)\mathcal{O}_{1,{-a_1/2}}\sigma_{2}(z_1) \mathcal{O}_{s,{-a_s/2}}\sigma_{2}(z_s)\mathcal{O}_{0,{-a_0/n}}\sigma_{n}(z_0)\rangle}{\langle \mathcal{O}_{\infty,{-a_\infty/n}}\sigma_{2}(z_\infty)\mathcal{O}_{1,{-a_1/2}}\sigma_{2}(z_1) \mathcal{O}_{s,{-a_s/2}}\sigma_{2}(z_s)\mathcal{O}_{0,{-a_0/n}}\sigma_{n}(z_0)\rangle} \nn \\
&= \left(\frac{(z_\infty-z_0)(z_1-z_\infty)}{(z_1-z_0)} \frac{(-1)s}{(n-1)s-n}\right)^{-2/n}\frac{1}{n} \nn \\
& \times \Bigg[  a_{\infty} \left(\frac{s^2(ns-n-s)^2-(2s^2-9)(s-1)^2-(5s+2)(s-1)^3(n-2)}{s^2(ns-n-s)^2}\right) \nn \\
&\quad + a_1 \left(\frac{2ns^2-2ns-5s^2+6s-3}{s(ns-n-s)}\right)\nn \\
&\quad + a_s \left(\frac{\splitfrac{-s^2(ns-n-s)^2+3(n^2-2)(s-1)^2+3(s-2)(s-1)^2(ns-n-s-2)}{-3(2n(s-1)-1)}}{s^2(ns-n-s)^2}\right)\nn \\
&\quad - a_0 \left(\frac{ns^2-ns-4s^2+6s-3}{s(ns-n-s)}\right)\nn \\
&\quad - \frac{f'}{f}\left(\frac{(ns-n-s-1)(ns-n-1)(ns^2-4s^2-n+6s-4)(s-1)}{s^3(ns-n-s)^3}\right) \nn \\
&\quad - \frac{c}{4}\frac{(s-1)^2((n-1)s^2-(n+1))}{((n-1)s-n)^2s^2}\Bigg]\nn
\end{align}
As before, all factors of $t_{ij}$ cancel.  Further, the factors of $z_{ij}$ above simply correct the explicit factors of $z_{ij}$ appearing in the 4-point function to be that of a four point function of primaries with appropriate conformal weights.  The sum over orbifold images my be accomplished by multiplying the last line above by the denominator of the first line and then summing over solutions to the polynomial $\zeta_z(s)$ (\ref{zetazdef}).  The remaining factor is a function of the map parameter $s$, which is the single parameter controlling the cross ratio $\zeta_z$.  Thus, the descendant correlator has been written in terms of the ancestor.

%\newpage

\section{Discussion}
\label{discsection}

In this work we have extended our previous work \cite{Burrington:2022dii} on fractional conformal descendants in correlators.  We have proceeded in two main directions.  First, we examine the case of three point functions with arbitrary size single cycle twists.  These are still in a class that are relatively easy to analyze, given that the $SL(2)$ symmetry on the base space allows us to fix the location of the operators, and the $SL(2)$ symmetry on the cover allows us to fix the locations of the ramified points in the map.  Using this, we find that all covering space information cancels out in our example calculations, and leaves descent relations which only depend on base space information in a very simple way, corroborating our earlier claims in \cite{Burrington:2022dii}.

In addition, we have explored higher $n$-point functions by considering an infinite class of 4-point functions.  While in the three-point functions we have seen explicitly that the covering space variables completely cancel out, with no residual need to sum over orbifold images.  In the four-point function, we see that the remaining parameter $s$ is exactly the one that controls both the cross ratio on the cover and on the base, but also controls which group product is being explored when OPE limits are taken (equally, considering which group product is being taken for a given choice of simply connected patch).  This, in fact, is as far as the 4-point functions have been explored here or elsewhere in the literature.

One may suspect that the sum over orbifold images is required at this stage to make the full answer a function only of the cross ratio $\zeta_z$.   There may be some direct proof of such statements using more sophisticated mathematical tools, e.g. Galois theory, or studies of algebraic varieties.  However, one may construct a more intuitive argument.  The answers presented here and elsewhere for 4-point functions depend on a single parameter $s$.  To solve for $s$ in terms of the cross ratio on the base $\zeta_z$ involves solving some polynomial equation in $s$ $P(s,\zeta_z)=0$ and so represents some algebraic curve (for the case considered in the main text text, see (\ref{zetaz}), where the polynomial is order $2n$).  The sum over orbifold images is presumably the sum over the all $s_i$ which solve this equation, i.e. the different branches of the algebraic curve.  Thus, after using covering space techniques to obtain the answer for a given $s_i$, one has a correlator of the correct form for a four point function multiplied by some functions $(f_L(s_i) f_R(s_i^*))$ where $f_L$ and $f_R$ are the individual parts coming from left and right movers (the complex conjugation makes these not just functions of the roots of $P(s)$, but rather the roots and complex conjugates).  Summing over values of $s_i$ which satisfy $P(s_i,\zeta_z)=0$ presumably renders the fully summed function $\sum_i (f_L(s_i)) f_R(s_i^*))$ dependent on the coefficients of the polynomial $P(s_i,\zeta_z)=0$, i.e. dependent on $\zeta$, rather than the roots themselves.  Interchange of the roots acts trivially on the summed form $\sum_i (f_L(s_i)) f_R(s_i^*))$, and also does not change the polynomial $P(s,\zeta)$ (the coefficients of $P(s,\zeta)$ are the elementary symmetric functions of the roots).  Thus, these summed functions appear to be only a function of the cross ratio in the base space $\zeta_z$ (and its complex conjugate).  However, this bears deeper scrutiny, even in the case of bare twists.

Finally, in higher point functions, it is also of interest to know how to control the location of the twisted operators.  This is particularly important in conformal perturbation theory where the locations of operators associated with deformations must be integrated over, and further must be regulated by applying infinitesimal size cutoffs around operators. Presumably the lifting maps have a list of ramified points and associated twist operators that can be controlled on both the base and the cover, although not independently.  The $SL(2)$ symmetries of the base and cover may fix the location of three operators on the base, and three ramified points on the cover.  The remaining degrees of freedom must be contained in a list of $n-3$ map parameters, which should specify both the $n-3$ independent cross ratios on the cover and $n-3$ cross ratios on the base.  Confluences of these simple cycle twist maps will give rise to lower $n$-point functions with compound group elements.  A full characterization of the link between map parameters and locations of twists would be interesting and helpful.

There is one more future direction which we are presently exploring: the superconformal case.  This expansion to considering other symmetry currents should allow us to address the deformation of the D1-D5 CFT.  At this point, we see very little difficulty in doing so.  In fact, the addition of supercurrents and R-symmetry currents should be markedly simpler to analyze.  The supercurrents and R-symmetry currents are conformal primaries, and so lift more easily to the covering surface than the stress tensor.  Furthermore, we have now analyzed in some detail the methods for using fractional currents in correlators.  Of course, there are the usual issues of dealing with both Ramond and Neveu-Schwarz sector fields which appear depending on the size of the twist \cite{Lunin:2001pw}.  These may be dealt with fairly directly using bosonization and the appropriate set of cocycles \cite{Burrington:2015mfa}.  We look forward to presenting such results, which we expect to be forthcoming.

\section*{Acknowledgements}

BAB and AWP wish to thank Ida G. Zadeh for feedback and suggestions on earlier versions of this work.  BAB is thankful for funding support from Hofstra Univeristy including startup funds and faculty research and development grants, and for support from the Scholars program at KITP, which is supported in part by the National Science Foundation under Grant No. NSF PHY-1748958.  The work of AWP is supported by a Discovery Grant from the Natural Sciences and Engineering Research Council of Canada.

%\vspace{8cm}

\appendix

\section{Jacobi Polynomials and the ($m$)-($n$)-($q$) fusion}
\label{JacobiSection}
\subsection{Collected identities for Jacobi Polynomials}
For more complete treatment of Jacobi Polynomials, see \cite{NIST} (this entry in the bibliography includes a link to an online version which is frequently updated).

The Jacobi polynomials are defined through the series expansion
\begin{align}
P_\gamma^{(\alpha,\beta)}(x)=\sum_{\ell=0}^\gamma \frac{(\gamma+\alpha+\beta+1)_\ell (\alpha+\ell+1)_{\gamma-\ell}}{\ell!(\gamma-\ell)!}\left(\frac{x-1}{2}\right)^\ell
\end{align}
where $(\kappa)_{\delta}=(\kappa)(\kappa+1)\times \cdots\times (\kappa+\delta-1)$ is Pochhammer's symbol (i.e. the ``rising factorial'' with $\delta$ terms).  This can be written in terms of the gamma function
\begin{equation}
(\kappa)_{\delta}=\frac{\Gamma(\kappa+\delta)}{\Gamma(\kappa)}
\end{equation}
keeping in mind that regulation may be necessary for negative integer values.  There is also the identity
\begin{equation}
(-\kappa)_{\delta}=(-1)^\delta (\kappa-\delta+1)_\delta.
\end{equation}

Some particularly useful values of Jacobi polynomials are (assuming $\gamma\geq 0$, and that $\alpha, \beta, \gamma$ are integers)
\begin{align}
P_\gamma^{(\alpha, \beta)}(1)= \frac{(\alpha+1)_\gamma}{\gamma!}= \begin{cases} \frac{(\alpha+\gamma)!}{\alpha! \gamma!}\qquad  &\text{if $\alpha\geq0$} \\
(-1)^\gamma \frac{ (-\alpha-1)!}{(-\alpha-\gamma-1)! \gamma!} \qquad &\text{if $\alpha\leq-1$ and $\alpha+\gamma+1\leq 0$} \label{valueat1}\\
0 \qquad &\text{otherwise}
\end{cases}
\end{align}

The Jacobi polynomials can also be written using the Rodrigues formula
\begin{equation}
P^{(\alpha,\beta)}_{\gamma}(x)= \frac{(-1)^\gamma}{2^\gamma \gamma!}(1-x)^{-\alpha}(1+x)^{-\beta} \left(\frac{d}{dx}\right)^\gamma\left((1-x)^{\alpha+\gamma}(1+x)^{\beta+\gamma}\right).
\end{equation}
The Jacobi polynomials have the following recursion relations and symmetries, most of which can be proved relatively quickly using either the series expression or Rodrigues formula above
\begin{align}
&P^{(\alpha,\beta)}_{\gamma}(-x)=(-1)^\gamma P^{\beta,\alpha}_{\gamma}(x) \\
&P^{\alpha,\beta-1}_{\gamma}(x)-P^{\alpha-1,\beta}_{\gamma}(x) = P^{(\alpha,\beta)}_{\gamma-1}(x) \\
&\frac{(1-x)}{2}P^{\alpha+1,\beta}_{\gamma}(x) +\frac{(1+x)}{2}P^{\alpha,\beta+1}_{\gamma}(x) =P^{(\alpha,\beta)}_{\gamma}(x) \\
&(2\gamma+\alpha+\beta+1)P^{(\alpha,\beta)}_{\gamma}(x) =(\gamma+\alpha+\beta+1)P^{(\alpha,\beta+1)}_{\gamma}(x) +(\gamma+\alpha)P^{(\alpha,\beta+1)}_{\gamma-1}(x) \\
&(2\gamma+\alpha+\beta+1)P^{(\alpha,\beta)}_{\gamma}(x) =(\gamma+\alpha+\beta+1)P^{(\alpha+1,\beta)}_{\gamma}(x) -(\gamma+\beta)P^{(\alpha+1,\beta)}_{\gamma-1}(x) \\
&(2\gamma+\alpha+\beta+2)\frac{(1+x)}{2}P^{(\alpha,\beta+1)}_{\gamma}(x)= (\gamma+1)P^{(\alpha,\beta)}_{\gamma+1}(x)+(\gamma+\beta+1)P^{(\alpha,\beta)}_{\gamma}(x) \\
&(2\gamma+\alpha+\beta+2)\frac{(1-x)}{2}P^{(\alpha+1,\beta)}_{\gamma}(x)= -(\gamma+1)P^{(\alpha,\beta)}_{\gamma+1}(x)+(\gamma+\alpha+1)P^{(\alpha,\beta)}_{\gamma}(x)
\end{align}
There is also a useful identity that connects Jacobi polynomials of different order, but equivalent indices:
\begin{align}
& P_{\gamma+1}^{(\alpha,\beta)}(x)=\left(A_{\gamma}^{(\alpha,\beta)} x + B_{\gamma}^{(\alpha,\beta)}\right)P_{\gamma}^{(\alpha,\beta)}(x) -C_{\gamma}^{(\alpha,\beta)}P_{\gamma-1}^{(\alpha,\beta)}(x) \\
& A_{\gamma}^{(\alpha,\beta)} = \frac{(2\gamma+\alpha+\beta+1)(2\gamma+\alpha+\beta+2)}{2(\gamma+1)(\gamma+\alpha+\beta+1)}\nn \\
& B_{\gamma}^{(\alpha,\beta)} = \frac{(\alpha^2-\beta^2)(2\gamma+\alpha+\beta+1)} {2(\gamma+1)(\gamma+\alpha+\beta+1)(2\gamma+\alpha+\beta)}\nn \\
& C_{\gamma}^{(\alpha,\beta)} = \frac{(\gamma+\alpha)(\gamma+\beta)(2\gamma+\alpha+\beta+2)} {(\gamma+1)(\gamma+\alpha+\beta+1)(2\gamma+\alpha+\beta)}\nn
\end{align}
In addition, the Jacobi Polynomials have the following first derivative
\begin{equation}
2\pa_x P^{(\alpha,\beta)}_{\gamma}(x)=(\gamma+\alpha+\beta+1)P^{(\alpha+1,\beta+1)}_{\gamma-1}(x)
\end{equation}
and satisfy the differential equation
\begin{align}
\pa_{x}^2 & P_{\gamma}^{(\alpha,\beta)}(x)+\left(\frac{\alpha+1}{x-1}+\frac{\beta+1}{x+1}\right)\pa_{x} P_{\gamma}^{(\alpha,\beta)}(x) \nn \\
& \qquad \qquad +\frac{\gamma(\gamma+\alpha+\beta+1)}{2}\left(\frac{-1}{x-1}+\frac{1}{x+1}\right)P_{\gamma}^{(\alpha,\beta)}(x)=0.
\end{align}

We adjust the formulae for the variable $x=2u-1$, which appears in the main text.  Here and elsewhere, when the argument is $x=2u-1$ we suppress the argument of the Jacobi polynomial to streamline notation.  Doing so, we find
\begin{align}
P_\gamma^{(\alpha,\beta)}=\sum_{\ell=0}^\gamma \frac{(\gamma+\alpha+\beta+1)_\ell (\alpha+\ell+1)_{\gamma-\ell}}{\ell!(\gamma-\ell)!}\left(u-1\right)^\ell, \label{seriesu}
\end{align}
\begin{equation}
P^{(\alpha,\beta)}_{\gamma}= \frac{1}{\gamma!}(u-1)^{-\alpha}(u)^{-\beta} \left(\frac{d}{du}\right)^\gamma\left((u-1)^{\alpha+\gamma}(u)^{\beta+\gamma}\right), \label{rodriguesu}
\end{equation}
and the recurrence relations become
\begin{align}
&P^{(\alpha,\beta)}_{\gamma}(1-2u)=(-1)^\gamma P^{\beta,\alpha}_{\gamma}(2u-1) \label{symu}\\
&P^{\alpha,\beta-1}_{\gamma}-P^{\alpha-1,\beta}_{\gamma} = P^{(\alpha,\beta)}_{\gamma-1} \label{id1} \\
&(1-u)P^{\alpha+1,\beta}_{\gamma} +uP^{\alpha,\beta+1}_{\gamma} =P^{(\alpha,\beta)}_{\gamma} \label{id2} \\
&(2\gamma+\alpha+\beta+1)P^{(\alpha,\beta)}_{\gamma} =(\gamma+\alpha+\beta+1)P^{(\alpha,\beta+1)}_{\gamma} +(\gamma+\alpha)P^{(\alpha,\beta+1)}_{\gamma-1} \label{id3}\\
&(2\gamma+\alpha+\beta+1)P^{(\alpha,\beta)}_{\gamma} =(\gamma+\alpha+\beta+1)P^{(\alpha+1,\beta)}_{\gamma} -(\gamma+\beta)P^{(\alpha+1,\beta)}_{\gamma-1} \label{id4} \\
&(2\gamma+\alpha+\beta+2)uP^{(\alpha,\beta+1)}_{\gamma}= (\gamma+1)P^{(\alpha,\beta)}_{\gamma+1}+(\gamma+\beta+1)P^{(\alpha,\beta)}_{\gamma} \label{id5} \\
&(2\gamma+\alpha+\beta+2)(1-u)P^{(\alpha+1,\beta)}_{\gamma}= -(\gamma+1)P^{(\alpha,\beta)}_{\gamma+1}+(\gamma+\alpha+1)P^{(\alpha,\beta)}_{\gamma} \label{id6}
\end{align}
The ``equal index'' recurrence formula remains unchanged (except for the variable substitution $x=2u-1$):
\begin{align}
& P_{\gamma+1}^{(\alpha,\beta)}=\left(A_{\gamma}^{(\alpha,\beta)}(2u-1) + B_{\gamma}^{(\alpha,\beta)}\right)P_{\gamma}^{(\alpha,\beta)} -C_{\gamma}^{(\alpha,\beta)}P_{\gamma-1}^{(\alpha,\beta)} \label{idorderonly} \\
& A_{\gamma}^{(\alpha,\beta)} = \frac{(2\gamma+\alpha+\beta+1)(2\gamma+\alpha+\beta+2)}{2(\gamma+1)(\gamma+\alpha+\beta+1)}\nn \\
& B_{\gamma}^{(\alpha,\beta)} = \frac{(\alpha^2-\beta^2)(2\gamma+\alpha+\beta+1)} {2(\gamma+1)(\gamma+\alpha+\beta+1)(2\gamma+\alpha+\beta)}\nn \\
& C_{\gamma}^{(\alpha,\beta)} = \frac{(\gamma+\alpha)(\gamma+\beta)(2\gamma+\alpha+\beta+2)} {(\gamma+1)(\gamma+\alpha+\beta+1)(2\gamma+\alpha+\beta)}.\nn
\end{align}
The derivative and differential equation become
\begin{equation}
\pa_u P^{(\alpha,\beta)}_{\gamma}=(\gamma+\alpha+\beta+1)P^{(\alpha+1,\beta+1)}_{\gamma-1} \label{derivu}
\end{equation}
\begin{align}
&\pa_{u}^2  P_{\gamma}^{(\alpha,\beta)}+\left(\frac{\alpha+1}{u-1}+\frac{\beta+1}{u}\right)\pa_{u} P_{\gamma}^{(\alpha,\beta)} -\frac{\gamma(\gamma+\alpha+\beta+1)}{u(u-1)}P_{\gamma}^{(\alpha,\beta)}=0. \label{diffequ}
\end{align}
We find it useful also to expand the Rodrigues formula
\begin{align}
P^{(\alpha,\beta)}_{\gamma}(2u-1)&= \frac{1}{\gamma!}(u-1)^{-\alpha}(u)^{-\beta} \left(\frac{d}{du}\right)^\gamma\left((u-1)^{\alpha+\gamma}(u)^{\beta+\gamma}\right) \nn \\
&=\frac{1}{\gamma!}(u-1)^{-\alpha}(u)^{-\beta} \sum_{\ell=0}^{\gamma} \binom{\gamma}{\ell}
\left[\left(\frac{d}{du}\right)^\ell (u-1)^{\alpha+\gamma}\right]\left[\left(\frac{d}{du}\right)^{(\gamma-\ell)}u^{\beta+\gamma}\right] \nn \\
&=\sum_{\ell=0}^{\gamma}\frac{(\alpha+\gamma-\ell+1)_{\ell}(\beta+\ell+1)_{\gamma-\ell}}{\ell!(\gamma-\ell)!} (u-1)^{\gamma-\ell}u^{\ell}
\label{rodriguesu2} \\
&=\sum_{\ell=0}^{\gamma}\frac{(\alpha+\ell+1)_{\gamma-\ell}(\beta+\gamma-\ell+1)_{\ell}}{\ell!(\gamma-\ell)!} (u-1)^{\ell}u^{\gamma-\ell}\label{rodriguesu22}
\end{align}
The expanded Rodrigues formula (\ref{rodriguesu22}) and (\ref{seriesu}) are easily related by comparing the series expansions about $u=1$, and using the relationship between generalized binomial coefficients
\begin{equation}
\frac{\Gamma(A+1)}{\Gamma(B+1)\Gamma(A-B+1)}=\sum_{\ell=0}^{\tilde{\ell}} \binom{\tilde{\ell}}{\ell}\frac{\Gamma(A-\tilde{\ell}+1)}{\Gamma(B-\ell+1)\Gamma(A-B-\tilde{\ell}+\ell+1)}.
\end{equation}

\subsection{Wronskian and Schwarzian for ($m$)-($n$)-($q$) Covering Maps}
\label{wronskandschwarz}
While this is a known result, it is a useful exercise prove the statement (\ref{dwdugen}) from the text.  First, directly taking the derivative of $w$, we find
\begin{align}
\pa_{u}w = \frac{(-1)^{(m-1)}u^{n-1}}{\left(P_{k-1}^{(-m,-n)}\right)^2} \left(nP^{(-m,n)}_{m-k}P_{k-1}^{(-m,-n)}+ u\pa_uP^{(-m,n)}_{m-k}P_{k-1}^{(-m,-n)}- uP^{(-m,n)}_{m-k}\pa_u P_{k-1}^{(-m,-n)}\right)
\end{align}
where we have suppressed the argument of all Jacobi polynomials, which are implied to be $2u-1$.  Plugging in the derivative (\ref{derivu}), we find
\begin{align}
& \pa_{u}w = \frac{(-1)^{(m-1)}u^{n-1}}{\left(P_{k-1}^{(-m,-n)}\right)^2} \Bigg(nP^{(-m,n)}_{m-k}P_{k-1}^{(-m,-n)} \nn \\
&\qquad \qquad +(n-k+1)uP^{(-m+1,n+1)}_{m-k-1}P_{k-1}^{(-m,-n)}+ (m+n-k)uP^{(-m,-n)}_{m-k}P_{k-2}^{(-m+1,-n+1)}\Bigg)
\end{align}
We may simplify the above relation by first eliminating explicit factors of $u$
\begin{align}
uP^{(-m+1,n+1)}_{m-k-1}& =u\left(P^{(-m+1,n)}_{m-k}-P^{(-m,n+1)}_{m-k}\right) \nn \\
&=-\left((1-u)P^{(-m+1,n)}_{m-k}+ uP^{(-m,n+1)}_{m-k}-P^{(-m+1,n)}_{m-k}\right)\nn \\
&=-(P^{(-m,n)}_{m-k}-P^{(-m+1,n)}_{m-k}) \label{removeu1}
\end{align}
where we have used relations (\ref{id1}) and then (\ref{id2}).  Using the exact same sequence of identities, one can show that
\begin{equation}
uP^{(-m+1,-n+1)}_{k-2}=-(P^{(-m,-n)}_{k-1}-P^{(-m+1,-n)}_{k-1}) \label{removeu2}
\end{equation}
One may be worried about the above identities when negative subscripts appear ($m=k$ in (\ref{removeu1}) and $k=1$ in (\ref{removeu2})).  In either case, $P^{(\alpha+1,\beta+1)}_{-1}$ should be interpreted as a derivative of $P^{(\alpha,\beta)}_{0}$.  However, $P^{(\alpha,\beta)}_0=1$, and so the derivative is 0, which also agrees with the right hand side of the above identities.

Plugging these in above, we find
\begin{align}
& \pa_{u}w = \frac{(-1)^{(m-1)}u^{n-1}}{\left(P_{k-1}^{(-m,-n)}\right)^2} \Bigg(-(n+m-2k+1)P^{(-m,n)}_{m-k}P_{k-1}^{(-m,-n)} \nn \\
&\qquad \qquad +(n-k+1)P^{(-m+1,n)}_{m-k}P_{k-1}^{(-m,-n)}+ (n+m-k)P^{(-m,n)}_{m-k}P_{k-1}^{(-m+1,-n)}\Bigg).
\end{align}
The first and third terms above combine using (\ref{id4}) with $\gamma=k-1,\alpha=-m,\beta=-n$, giving
\begin{align}
& \pa_{u}w = \frac{(-1)^{(m-1)}u^{n-1}}{\left(P_{k-1}^{(-m,-n)}\right)^2} (n-k+1)\Bigg(P^{(-m,n)}_{m-k}P_{k-2}^{(-m+1,-n)} + P^{(-m+1,n)}_{m-k}P_{k-1}^{(-m,-n)}\Bigg).
\end{align}

We now directly contend with the factor
\begin{align}
& Q_k=(n-k+1)\Bigg(P^{(-m,n)}_{m-k}P_{k-2}^{(-m+1,-n)} + P^{(-m+1,n)}_{m-k}P_{k-1}^{(-m,-n)}\Bigg)
\end{align}
First, we note that the above form is simple for $k=1$ (keeping in mind that Jacobi polynomials with negative subscripts are 0, and $P^{(\alpha,\beta)}_{0}=1$), we see that
\begin{align}
Q_1=nP^{(-m+1,n)}_{m-1}
\end{align}
which we can evaluate using the Rodrigues formula (\ref{rodriguesu})
\begin{align}
nP^{(-m+1,n)}_{m-1}& =n\frac{1}{(m-1)!} (u-1)^{m-1} u^{-n} \frac{d^{m-1}}{du^{m-1}}\left((u-1)^{0}u^{n+m-1}\right) \nn \\
&=(u-1)^{m-1}\frac{(n+m-1)!}{(n-1)!(m-1)!}.
\end{align}
which agrees with (\ref{dwdugen}) with $k=1$.  Another way of seeing this is by using the series representation of $P_{\gamma}^{(-\gamma,\beta)}$, (\ref{seriesu}), and noticing that  the Pochhammer symbol $(-\gamma+\ell+1)_{\ell-\gamma}$ vanishes except when $\ell=\gamma$.

Thus, all that remains to do is to show that
\begin{equation}
Q_{k+1}=-\frac{(m-k)(n-k)}{(n+m-k)(k)} Q_k. \label{neededrecur}
\end{equation}
Obviously this induction only connects $Q_k$ for the allowed values $k\geq1$, $k\leq m$, $k\leq n$.  We start with
\begin{equation}
Q_{k+1}=(n-k)\left(P^{(-m,n)}_{m-k-1}P^{(-m+1,-n)}_{k-1} +P^{(-m+1,n)}_{m-k-1}P^{(-m,-n)}_{k}\right)\label{Qkp1start}
\end{equation}
To help us organize terms, we recognize that there are certain functions of $\alpha$, $\beta$, and $\gamma$ that come up in the recurrence relations, and there are two basic families of Jacobi polynomials above.  \\
\begin{center}
\begin{tabular}{c|c|c|c|c|c|}
%\hline
& $2\gamma+\alpha+\beta+1$ & $\gamma+\alpha+\beta+1$ & $\gamma+1$ & $\gamma+\alpha$ & $\gamma+\beta$ \\
\hline
$P^{(-m,n)}_{m-k}$ & $m+n-2k+1$ & $n-k+1$ & $m-k+1$ & $-k$ & $m+n-k$ \\
\hline
$P^{(-m,-n)}_{k-1}$ & $-m-n+2k-1$ & $-m-n+k$ & $k$ & $-m+k-1$ & $-n+k-1$ \\
\hline
\end{tabular}
\end{center}
Every entry in the first row appears in the second row with an additional minus sign, and coming from a different combination of the parameters $\alpha$, $\beta$, and $\gamma$.  Of course in (\ref{Qkp1start}) the factors appear shifted by small integers, however, the above table helps us identify the necessary steps to introduce the factors appearing in (\ref{neededrecur}).

First, we use (\ref{id6}) with $\alpha=-m$, $\beta=-n$, $\gamma=k-1$ in the second term in (\ref{Qkp1start}) to find
\begin{align}
Q_{k+1}&=(n-k)\Bigg(P^{(-m,n)}_{m-k-1}P^{(-m+1,-n)}_{k-1} \nn \\
&\qquad \qquad + P^{(-m+1,-n)}_{m-k-1}\frac{1}{k}\left[-(m-k)P^{(-m,-n)}_{k-1} +(m+n-2k)(1-u)P^{(-m+1,-n)}_{k-1}\right]\Bigg) \nn \\
&=(n-k)\Bigg(P^{(-m,n)}_{m-k-1}P^{(-m+1,-n)}_{k-1} -\frac{(m-k)}{k}P^{(-m+1,n)}_{m-k-1}P^{(-m,-n)}_{k-1} \nn \\
&\qquad \qquad \qquad \qquad +\frac{(m+n-2k)}{k}(1-u)P^{(-m+1,-n)}_{m-k-1}P^{(-m+1,-n)}_{k-1}\Bigg)
\end{align}
The last term above suggests the use of (\ref{id6}) with $\gamma=m-k-1$, $\alpha=-m$, $\beta=-n$ (and so we have used identity (\ref{id6}) twice to ``pull a $(1-u)$ over'') resulting in
\begin{align}
Q_{k+1}&=(n-k)\Bigg(P^{(-m,n)}_{m-k-1}P^{(-m+1,-n)}_{k-1} -\frac{(m-k)}{k}P^{(-m+1,n)}_{m-k-1}P^{(-m,-n)}_{k-1} \nn \\
&\qquad \qquad \qquad \qquad +\frac{1}{k}\left[-(m-k)P^{(-m,n)}_{m-k}-kP^{(-m,n)}_{m-k-1}\right]P^{(-m+1,-n)}_{k-1}\Bigg) \nn \\
&=-\frac{(n-k)(m-k)}{k}\Bigg(P^{(-m+1,n)}_{m-k-1}P^{(-m,-n)}_{k-1} +P^{(-m,n)}_{m-k}P^{(-m+1,-n)}_{k-1}\Bigg)
\end{align}
The above combination is beginning to look correct, although the $-m+1$ are on the incorrect terms.  This suggests the use of (\ref{id4}) on the first term with $\gamma=m-k$, $\alpha=-m$, $\beta=n$, finding
\begin{align}
Q_{k+1}
&= -\frac{(n-k)(m-k)}{k}\Bigg(\frac{(n-k+1)}{(n+m-k)}P^{(-m+1,n)}_{m-k}P^{(-m,-n)}_{k-1} \nn \\
& \qquad \qquad -\frac{(n+m-2k+1)}{(n+m-k)} P^{(-m,n)}_{m-k}P^{(-m,-n)}_{k-1} +P^{(-m,n)}_{m-k}P^{(-m+1,-n)}_{k-1}\Bigg)
\end{align}
and now we may again use (\ref{id4}) on the second term above with $\gamma=k-1$, $\alpha=-m$, $\beta=-n$ to reabsorb the factor of $-(n+m-2k+1)$ into the second Jacobi polynomial, giving
\begin{align}
Q_{k+1}
&= -\frac{(n-k)(m-k)}{k}\Bigg(\frac{(n-k+1)}{(n+m-k)}P^{(-m+1,n)}_{m-k}P^{(-m,-n)}_{k-1} \nn \\
& \qquad \qquad +\frac{1}{(n+m-k)} P^{(-m,n)}_{m-k}\left(-(n+m-k)P^{(-m+1,-n)}_{k-1}+(n-k+1)P^{(-m+1,-n)}_{k-2}\right) \nn \\
&\qquad \qquad \qquad+P^{(-m,n)}_{m-k}P^{(-m+1,-n)}_{k-1}\Bigg)
\end{align}
and we get a cancellation between terms, finally giving
\begin{align}
Q_{k+1}
&= -\frac{(n-k)(m-k)}{(n+m-k)(k)}\Bigg((n-k+1)\left(P^{(-m+1,n)}_{m-k}P^{(-m,-n)}_{k-1} +P^{(-m,n)}_{m-k}P^{(-m+1,-n)}_{k-2}\right) \Bigg) \nn \\
&= -\frac{(n-k)(m-k)}{(n+m-k)(k)}Q_k.
\end{align}
This concludes the proof of (\ref{dwdugen}) in the main text, i.e.
\begin{equation}
\pa_{u} w= \frac{1}{\left(P^{(-m,-n)}_{k-1}(2u-1)\right)^2} u^{n-1}(u-1)^{m-1}\frac{(-1)^{m-k}(n+m-k)!}{(k-1)!(m-k)!(n-k)!}
\label{dwdugenequationapx}
\end{equation}

We now turn our attention to proving (\ref{schwarzgen}) in the main text.  This may be computed more quickly, given the discussion of the Schwarzian in other places, for example \cite{HilleODE}.  However, for completeness, we show all details here.

First, there are many ways of writing the Schwarzian derivative.  However, we find a particularly convenient way as
\begin{equation}
\left\{w,u\right\}=\pa_u\left(\pa_u\ln(\pa_u w)\right)-\frac{1}{2}\left(\pa_u\ln(\pa_u w)\right)^2 \label{schwarzlog}
\end{equation}
which immediately takes advantage of the scaling independence of the Schwarzian. First, we compute
\begin{equation}
\pa_u\ln(\pa_u w)=\frac{(n-1)}{u}+\frac{(m-1)}{(u-1)}-2\frac{\pa_u P^{(-m,-n)}_{k-1}}{P^{(-m,-n)}_{k-1}}
\end{equation}
and
\begin{equation}
\pa_u(\pa_u\ln(\pa_u w))=-\frac{(n-1)}{u^2}-\frac{(m-1)}{(u-1)^2}-2\frac{\pa_u^2 P^{(-m,-n)}_{k-1}}{P^{(-m,-n)}_{k-1}}+2\left(\frac{\pa_u P^{(-m,-n)}_{k-1}}{P^{(-m,-n)}_{k-1}}\right)^2
\end{equation}
which we plug into (\ref{schwarzlog}) and find
\begin{align}
\left\{w,u\right\}& =-\frac{n^2-1}{2u^2}-\frac{m^2-1}{2(u-1)^2}-\frac{(n-1)(m-1)}{u(u-1)} \nn \\
&-\frac{2}{P^{(-m,-n)}_{k-1}}\left(\pa_u^2P^{(-m,-n)}_{k-1} +\left(\frac{(-n+1)}{u}+\frac{(-m+1)}{u-1}\right)\pa_uP^{(-m,-n)}_{k-1} \right).
\end{align}
In the last term, involving the Jacobi polynomials, we recognize the first two terms of the differential equation (\ref{diffequ}), and so can be replaced with the third term in the differential equation, causing the $P^{(-m,-n)}_{k-1}$ to cancel, yeilding
\begin{align}
\left\{w,u\right\}& =-\frac{n^2-1}{2u^2}-\frac{m^2-1}{2(u-1)^2}-\frac{(n-1)(m-1)+2(k-1)(k-m-n)}{u(u-1)} \nn\\
&=-\frac{n^2-1}{2u^2}-\frac{m^2-1}{2(u-1)^2}+\frac{n^2+m^2-q^2-1}{2u(u-1)}
\end{align}
where we have substituted in $k=(n+m-q+1)/2$.  Finally, we may use (\ref{utmnq}) and construct
\begin{align}
&\left\{z,t\right\}=\left\{w,u\right\}\left(\frac{du}{dz}\right)^2 \nn \\
&=-\frac{n^2-1}{2(t-t_n)^2}-\frac{m^2-1}{2(t-t_m)^2}-\frac{q^2-1}{2(t-t_q)^2} \nn \\
&\qquad  +\frac{(n^2-1)+(m^2-1)-(q^2-1)}{2(t-t_n)(t-t_m)} +\frac{(n^2-1)+(q^2-1)-(m^2-1)}{2(t-t_n)(t-t_q)} \nn \\
&\qquad \qquad +\frac{(m^2-1)+(q^2-1)-(n^2-1)}{2(t-t_m)(t-t_q)}
\end{align}
matching (\ref{schwarzgen}) in the main text.

\subsection{$m,n,q$ Exchange Symmetry for ($m$)-($n$)-($q$) Covering Maps}
\label{mnqsym}

In this subsection we establish the $m,n,q$ exchange symmetry of the three point single cycle covering space map $z(w(u(t)))$ given by
\begin{align}
& z(w)=\frac{z_{q}(z_m-z_{n})w-z_n(z_m-z_{q})} {(z_m-z_{n})w-(z_m-z_{q})} \nn \\
& w(u)=(-1)^{m-1} u^n \frac{P_{m-k}^{(-m,n)}(2u-1)}{P_{k-1}^{(-m,-n)}(2u-1)} \label{ztfullapx}  \\
& u(t)=\frac{t_m-t_{q}}{t_m-t_n}\frac{t-t_n}{t-t_{q}}  \nn
\end{align}
The symmetry ultimately comes from the underlying transformation properties of the Jacobi polynomials under $sl(2)$ transformations, however, we find it more enlightening to work the details out directly.

We start first by showing the above is $m\leftrightarrow n$ symmetric.  Proceeding layer by layer, we see
\begin{align}
[m\leftrightarrow n] z(w)=z(1-[m\leftrightarrow n] w).
\end{align}
To continue, we note that
\begin{align}
[m\leftrightarrow n] u(t)=(1-u(t)).
\end{align}
Thus, the argument of the $z$ function becomes
\begin{equation}
1-[m\leftrightarrow n] w=1-\frac{(-1)^{n-1}(1-u)^mP^{(-n,m)}_{n-k}(1-2u)}{P^{-n,-m}_{k-1}(1-2u)}.
\end{equation}
Using the identity (\ref{symu}) and suppressing the subsequent arguments of Jacobi functions (as usual, once the arguments are $2u-1$), we find
\begin{equation}
1-[m\leftrightarrow n] w=1+\frac{(-1)^{m-1}(u-1)^mP^{(m,-n)}_{n-k}}{P^{-m,-n}_{k-1}}
\end{equation}
Thus, the map is $m\leftrightarrow n$ symmetric if
\begin{equation}
1+\frac{(-1)^{m-1}(u-1)^mP^{(m,-n)}_{n-k}}{P^{-m,-n}_{k-1}}=\frac{(-1)^{m-1}u^nP^{(-m,n)}_{m-k}}{P^{-m,-n}_{k-1}}
\end{equation}
i.e. if $(1-[m\leftrightarrow n] w)=w$, which we set about proving now.  First, we rearrange this formula to
\begin{equation}
P^{(-m,-n)}_{k-1}=(-1)^{m-1}\left(u^nP^{(-m,n)}_{m-k}-(u-1)^mP^{(m,-n)}_{n-k}\right) \label{provemnsym}
\end{equation}
which we note is linear in Jacobi polynomials, suggesting an inductive proof.  We start with a trivial case $k=1$ $n=1$ to start an inductive proof \footnote{which on the face of it seems too simple: a twist $n=1$ is not a twist at all.  However, one can start with $k=1,n=2$, and show (\ref{provemnsym}) is correct in this case as well, which we have done, but omit here for brevity.}.  The above becomes
\begin{align}
& P^{(-m,-1)}_{0}=(-1)^{m-1}\left(uP^{(-m,1)}_{m-1}-(u-1)^mP^{(m,-1)}_{0}\right) \nn \\
& 1=(-1)^{m-1}\left(uP^{(-m,1)}_{m-1}-(u-1)^m\right) \label{Pk1n1}
\end{align}
Using the series representation, we find
\begin{equation}
P^{(-m,1)}_{m-1}=\sum_{\ell=0}^{m-1}\frac{(1)_{\ell}(-m+\ell+1)_{m-1-\ell}}{\ell!(m-1-\ell)!}(u-1)^\ell
\end{equation}
above, the Pochhammer symbols may be replaced with ordinary factorials, giving an incomplete geometric series
\begin{align}
P^{(-m,1)}_{m-1}& =(-1)^{m-1}\sum_{\ell=0}^{m-1}(1-u)^\ell  \nn \\
& =(-1)^{m-1}\left(\sum_{\ell=0}^{\infty}(1-u)^\ell-(1-u)^m\sum_{\ell=0}^{\infty}(1-u)^\ell\right)\
&=(-1)^{m-1}\frac{1-(1-u)^m}{u}
\end{align}
Plugging this into (\ref{Pk1n1}) concludes the first step in the induction.  Next, we assume that we have shown all $k=1$ cases up to some particular $n-1$, i.e.
\begin{align}
P^{-m,-n+1}_{0}=1=(-1)^{m-1}\left(u^{n-1}P^{(-m,n-1)}_{m-1}-(u-1)^mP^{(m,-n+1)}_{n-2}\right)
\end{align}
and proceed to show that the $n$ case follows from the $n-1$ case.  First, we replace $P^{(-m,n-1)}_{m-1}=uP^{(-m,n)}_{m-1}-(u-1)P^{(-m+1,n-1)}_{m-1}$ using (\ref{id2}), and replace $P^{(m,-n+1)}_{n-2}=P^{(m,-n)}_{n-1}-P^{(m-1,-n+1)}_{n-1}$ using (\ref{id1}), finding
\begin{align}
1&=(-1)^{m-1}\left(u^{n}P^{(-m,n)}_{m-1}-(u-1)^mP^{(m,-n)}_{n-1}\right)  \nn \\
& \qquad \qquad \qquad \qquad +(-1)^{m}\left(u^{n-1}(u-1)P^{(-m+1,n-1)}_{m-1}-(u-1)^mP^{(m-1,-n+1)}_{n-1}\right).
\end{align}
The second line above can be easily shown to be 0 by using the Rodrigues formula on each term
\begin{align}
(u-1)^mP^{(m-1,-n+1)}_{n-1} &=(u-1)^m\frac{1}{(n-1)!}(u-1)^{-m+1}u^{n-1}(\pa_u)^{n-1}\left((u-1)^{m+n-2}u^0\right) \nn \\
&= (u-1)^m u^{n-1}\frac{(m+n-2)!}{(m-1)!(n-1)!}  \label{rodsp1}\\
u^{n-1}(u-1)P^{(-m+1,n-1)}_{m-1} & = u^{n-1}(u-1)\frac{1}{(m-1)!} (u-1)^{m-1}u^{-n+1}\left(\frac{d}{du}\right)^{m-1}\left((u-1)^{0}u^{m+n-2}\right) \nn \\
&= (u-1)^m u^{n-1} \frac{(m+n-2)!}{(m-1)!(n-1)!} \label{rodsp2}
\end{align}
Thus, we have that
\begin{equation}
(-1)^{m-1}\left(u^{n}P^{(-m,n)}_{m-1}-(u-1)^mP^{(m,-n)}_{n-2}\right)=1=P_0^{-m,-n}
\end{equation}
because all Jacobi polynomials of degree 0 are equal to 1.  Thus, we have now shown (\ref{provemnsym}) generally for the $k=1$ case.  To prove (\ref{provemnsym}), we proceed with a proof by induction on $k$, which shifts the orders of Jacobi polynomials on both sides, suggesting the use of (\ref{idorderonly}).  However, we note that (\ref{idorderonly}) relates orders that differ by 2, not one, and so one would normally have to prove the $k=1$ case separately.  However, plugging in $\gamma=0$ into (\ref{idorderonly}), we find
\begin{equation}
P_{1}^{(\alpha,\beta)}=\left(A_{0}^{(\alpha,\beta)}(2u-1) + B_{0}^{(\alpha,\beta)}\right)P_{0}^{(\alpha,\beta)} -C_{0}^{(\alpha,\beta)}P_{-1}^{(\alpha,\beta)}
\end{equation}
which remains true if we interpret $P_{-1}^{(\alpha,\beta)}=0$.  This suggests proving (\ref{provemnsym}) for the special case $k=0$ as well, interpreting $P_{-1}^{(\alpha,\beta)}=0$.  Thus, we show that
\begin{align}
P_{-1}^{-m,-n}=0=(-1)^m\left(u^nP_{m}^{(-m,n)}-(u-1)^mP_{n}^{(m,-n)}\right),
\end{align}
however this has already been proven in (\ref{rodsp1}) and (\ref{rodsp2}) replacing $m-1\rightarrow m, n-1\rightarrow n$.

Thus, we have shown (\ref{provemnsym}) for the special cases $k=0$ and $k=1$, leaving us free to use and inductive proof on $k$ via (\ref{idorderonly}).  First, we assume that (\ref{provemnsym}) has been proven up to a given $k+1$, and show that the $k+2$ case follows. Using (\ref{idorderonly})
\begin{equation}
P_{k+1}^{-m,-n}=(A_{k}^{(-m,-n)}(2u-1)+B_{k}^{(-m,-n)})P^{(-m,-n)}_{k}-C_k^{(-m,-n)}P^{(-m,-n)}_{k-1}
\end{equation}
and then using the inductive assumption gives
\begin{align}
P_{k+1}^{-m,-n}&=(-1)^{m-1}\Bigg[
u^n\left((A_{k}^{(-m,-n)}(2u-1)+B_{k}^{(-m,-n)})P^{(-m,n)}_{m-k-1}-C_k^{(-m,-n)}P^{(-m,n)}_{m-k}\right) \nn \\
& \qquad - (u-1)^m\left((A_{k}^{(-m,-n)}(2u-1)+B_{k}^{(-m,-n)})P^{(m,-n)}_{n-k-1}-C_k^{(-m,-n)}P^{(m,-n)}_{n-k}\right)
\end{align}
Now, we will conclude our proof if it is the case that both of the following equations hold:
\begin{align}
&(A_{k}^{(-m,-n)}(2u-1)+B_{k}^{(-m,-n)})P^{(-m,n)}_{m-k-1}-C_k^{(-m,-n)}P^{(-m,n)}_{m-k}=P^{(-m,n)}_{m-k-2} \nn \\
&(A_{k}^{(-m,-n)}(2u-1)+B_{k}^{(-m,-n)})P^{(m,-n)}_{n-k-1}-C_k^{(-m,-n)}P^{(m,-n)}_{n-k}=P^{(m,-n)}_{n-k-2}
\end{align}
We see that these have a resemblance to (\ref{idorderonly}).  Rearranging, the above equations according to the order of the Jacobi polynomials, we find the above hold only if
\begin{align}
&P^{(-m,n)}_{(m-k-1)+1}=\left(\frac{A_{k}^{(-m,-n)}}{C_k^{(-m,-n)}}(2u-1)+\frac{B_{k}^{(-m,-n)}}{C_k^{(-m,-n)}}\right)P^{(-m,n)}_{m-k-1}-\frac{1}{C_k^{(-m,-n)}}P^{(m,-n)}_{n-k-2} \nn \\
&P^{(m,-n)}_{(n-k-1)+1}=\left(\frac{A_{k}^{(-m,-n)}}{C_k^{(-m,-n)}}(2u-1)+\frac{B_{k}^{(-m,-n)}}{C_k^{(-m,-n)}}\right)P^{(m,-n)}_{n-k-1}-\frac{1}{C_k^{(-m,-n)}}P^{(m,-n)}_{n-k-2}
\end{align}
These indeed hold because, checking explicitly
\begin{align}
&A_{m-k-1}^{(-m,n)}=\frac{A_{k}^{(-m,-n)}}{C_k^{(-m,-n)}}, \qquad B_{m-k-1}^{(-m,n)}=\frac{B_{k}^{(-m,-n)}}{C_k^{(-m,-n)}}, \qquad \qquad C_{m-k-1}^{(-m,n)}=\frac{1}{C_k^{(-m,-n)}}, \nn\\
&A_{n-k-1}^{(m,-n)}=\frac{A_{k}^{(-m,-n)}}{C_k^{(-m,-n)}}, \qquad B_{n-k-1}^{(m,-n)}=\frac{B_{k}^{(-m,-n)}}{C_k^{(-m,-n)}}, \qquad \qquad C_{n-k-1}^{(m,-n)}=\frac{1}{C_k^{(-m,-n)}},
\end{align}

Thus, we have proven (\ref{provemnsym}), and so the map is $m\leftrightarrow n$ interchange symmetric for all $m\geq k, n\geq k$, $m\geq 1, n\geq1$, and $k\geq 0$ which are the cases of relevance for the twist maps.

We proceed to examine the $n\leftrightarrow q$ interchange.  We again proceed in steps, noting that
\begin{align}
[n\leftrightarrow q] z(w)=z\left(\frac{1}{[n\leftrightarrow q]w}\right)
\end{align}
and
\begin{align}
[n\leftrightarrow q] u(t)=\frac{1}{u}.
\end{align}
The exchange $[n\leftrightarrow q]$ acting on the Jacobi polynomials must be taken with a bit of care.  While the exchange $[m\leftrightarrow n]$ leaves the overlap number $k=k_{m,n}$ invariant, this is no longer the case when switching $[n\leftrightarrow q]$.  Thus, we prefer to give the $w(u)$ function in its ``bare'' form, with all $n,m,q$ explicitly written, leaving the overlaps implied through the relation $k_{m,n}=\frac{1}{2}(m+n+1-q)$ and the $m,n,q$ permutations (with $k_{i,j}=k_{j,i}$),
\begin{equation}
w(u)=(-1)^{m-1} u^n \frac{P_{(m+q-n-1)/2}^{(-m,n)}(2u-1)}{P_{(m+n-q-1)/2}^{(-m,-n)}(2u-1)}.
\end{equation}
Thus, we find
\begin{equation}
\frac{1}{[n\leftrightarrow q]w}=(-1)^{m-1}u^q\frac{P_{(m+q-n-1)/2}^{(-m,-q)}(2/u-1)}{P_{(m+n-q-1)/2}^{(-m,q)}(2/u-1)}
\end{equation}
Therefore, if this is equal to $w$, the map remains invariant, and so we wish to prove
\begin{equation}
u^q\frac{P_{(m+q-n-1)/2}^{(-m,-q)}(2/u-1)}{P_{(m+n-q-1)/2}^{(-m,q)}(2/u-1)}= u^n \frac{P_{(m+q-n-1)/2}^{(-m,n)}(2u-1)}{P_{(m+n-q-1)/2}^{(-m,-n)}(2u-1)} \label{provenq}
\end{equation}
We first tackle the change in argument of the Jacobi polynomails using the expanded Rodrigues formula (\ref{rodriguesu2}), substituting $u\rightarrow 1/u$
\begin{align}
P^{(\alpha,\beta)}_{\gamma}(2/u-1)&= \sum_{\ell=0}^{\gamma}\frac{(\alpha+\gamma-\ell+1)_{\ell}(\beta+\ell+1)_{\gamma-\ell}}{\ell!(\gamma-\ell)!} (1/u-1)^{\gamma-\ell}1/u^{\ell} \nn \\
&=\frac{1}{u^\gamma}\sum_{\ell=0}^{\gamma}\frac{(\alpha+\gamma-\ell+1)_{\ell}(\beta+\ell+1)_{\gamma-\ell}} {\ell!(\gamma-\ell)!} (1-u)^{\gamma-\ell} \nn \\
&=\frac{1}{u^\gamma}\sum_{\ell=0}^{\gamma}\frac{(\alpha+\ell+1)_{\gamma-\ell}(\beta+\gamma-\ell+1)_{\ell}} {\ell!(\gamma-\ell)!}(-1)^\ell(u-1)^{\ell} \nn \\
&=\frac{1}{u^\gamma}\sum_{\ell=0}^{\gamma}\frac{(\alpha+\ell+1)_{\gamma-\ell}(-\beta-\gamma)_{\ell}} {\ell!(\gamma-\ell)!}(u-1)^{\ell}
\end{align}
where in the last step we have pushed the $(-1)^\ell$ inside the Pochhammer symbol of order $\ell$.  We recognize this final sum as being a Jacobi polynomial in the form (\ref{seriesu}), and so we find
\begin{equation}
P^{(\alpha,\beta)}_{\gamma}(2/u-1)=\frac{1}{u^\gamma} P^{(\alpha,-\beta-\alpha-2\gamma-1)}_\gamma(2u-1).
\end{equation}
Using the above formula in the two Jacobi functions on the left hand side of (\ref{provenq}) makes the left hand side become precisely the right hand side.  Thus, the covering map is also $[n \leftrightarrow q]$ symmetric as well.  Thus, the covering space map is fully $m,n,q$ interchange symmetric.

In the course of these proofs, we have encountered useful forms of the map $w(u)$ that make the ramification of the map manifest at the different points:
\begin{align}
w(u)&=u^n
\frac{P^{(n,-m)}_{\frac{1}{2}(m+q-n-1)}(1-2u)}{P^{-n,-m}_{\frac{1}{2}(m+n-q-1)}(1-2u)} \\
(1-w(u))&=(1-u)^m\frac{P^{(m,-n)}_{\frac{1}{2}(n+q-m-1)}(2u-1)}{P^{-m,-n}_{\frac{1}{2}(m+n-q-1)}(2u-1)} \\
\frac{1}{w(u)}&= \left(\frac{1}{u}\right)^q \frac{P^{(q,-m)}_{\frac{1}{2}(m+n-q-1)}(1-2/u)}{P^{(-q,-m)}_{\frac{1}{2}(m+q-n-1)}(1-2/u)}
\end{align}
which are useful in the $(t \rightarrow t_n, u\rightarrow 0)$,  $(t \rightarrow t_m, u\rightarrow 1)$, and $(t \rightarrow t_q, u\rightarrow \infty)$ limits respectively.

\section{Expansion polynomials for exampls 3 and 4 point functions}
\label{appxpoly}
\subsection{Expansion polynomials for $(m)-(n)-(q)$ 3-point function}
In the text, the polynomials in $m,n,q$ and $t_m,t_n,t_q$ are found to be
\begin{align}
Q_{t,n,2}(m,n,q)&=n\bigg( 2 m^4 n t_m^2-4 m^4 n t_m t_q+2 m^4 n t_q^2-4 m^2 n^3 t_m^2\nn\\
&+8 m^2 n^3 t_m t_n-8 m^2 n^3 t_n t_q+4 m^2 n^3 t_q^2-4 m^2 n q^2 t_m^2+8 m^2 n q^2 t_m t_q \nn \\
&-4 m^2 n q^2 t_q^2+2 n^5 t_m^2-8 n^5 t_m t_n+4 n^5 t_m t_q+8 n^5 t_n^2-8 n^5 t_n t_q \nn \\
&+2 n^5 t_q^2+4 n^3 q^2 t_m^2-8 n^3 q^2 t_m t_n+8 n^3 q^2 t_n t_q-4 n^3 q^2 t_q^2+2 n q^4 t_m^2\\
&-4 n q^4 t_m t_q+2 n q^4 t_q^2-5 m^4 t_m^2+10 m^4 t_m t_q-5 m^4 t_q^2+6 m^2 n^2 t_m^2\nn \\
&-8 m^2 n^2 t_m t_n-4 m^2 n^2 t_m t_q+8 m^2 n^2 t_n t_q-2 m^2 n^2 t_q^2+10 m^2 q^2 t_m^2\nn \\
&-20 m^2 q^2 t_m t_q+10 m^2 q^2 t_q^2-n^4 t_m^2+8 n^4 t_m t_n-6 n^4 t_m t_q-8 n^4 t_n^2+8 n^4 t_n t_q \nn \\
&-n^4 t_q^2-2 n^2 q^2 t_m^2+8 n^2 q^2 t_m t_n-4 n^2 q^2 t_m t_q-8 n^2 q^2 t_n t_q+6 n^2 q^2 t_q^2\nn \\
&-5 q^4 t_m^2+10 q^4 t_m t_q-5 q^4 t_q^2+12 m^2 n t_m^2-32 m^2 n
t_m t_n+8 m^2 n t_m t_q \nn \\ &+32 m^2 n t_n t_q-20 m^2 n t_q^2-12 n^3 t_m^2+40 n^3 t_m t_n-16 n^3 t_m t_q-40 n^3 t_n^2 \nn \\
&+40 n^3 t_n t_q-12 n^3
t_q^2-20 n q^2 t_m^2+32 n q^2 t_m t_n+8 n q^2 t_m t_q-32 n q^2 t_n t_q \nn \\
&+12 n q^2 t_q^2-14 m^2 t_m^2+32 m^2 t_m t_n-4 m^2 t_m t_q-32 m^2 t_n t_q+18 m^2 t_q^2+6 n^2 t_m^2 \nn \\
&-40 n^2 t_m t_n+28 n^2 t_m t_q+40 n^2 t_n^2-40 n^2 t_n t_q+6 n^2 t_q^2+18 q^2 t_m^2-32 q^2 t_m t_n\nn \\
&-4 q^2 t_m t_q+32 q^2 t_n t_q-14 q^2 t_q^2+18 n t_m^2-32 n t_m t_n-4 n t_m t_q+32 n t_n^2\nn \\
&-32 n t_n t_q+18 n t_q^2-13 t_m^2+32 t_m t_n-6 t_m t_q-32 t_n^2+32 t_n t_q-13 t_q^2
\bigg)
\end{align}
The polynomial $Q_{t,n,3}(m,n,q)$ is not particularly illuminating, and we omit it here.

\subsection{Expansion polynomials for $(n)-(2)-(2)-(n)$ 4-point function}
\label{expansionAppx}
\begin{align}
&S_{1,0}=-\frac{n\big((ns-(n+1))t_{1,0}-(s-1)^2t_{1,\infty}\big)}{t_{1,0}t_{0,\infty}(ns-(n+1))} \\
&S_{2,0}=\frac{n\left((n+1)(ns-(n+1))t_{1,0}^2-2(n+1)(s-1)^2t_{1,0}t_{1,\infty}+2s(s-1)^2t_{1,\infty}^2\right)}{2t_{1,0}^2t_{0,\infty}^2(ns-(n+1))}
\end{align}
\begin{align}
&S_{1,\infty}=\frac{n\left(s((n-1)s-n)t_{1,\infty}+(s-1)^2t_{0,1}\right)}{t_{1,\infty}t_{0,\infty} s((n-1)s-n)}  \\
&S_{2,\infty}=\frac{n\Bigg(\splitdfrac{2(s-1)^2(ns-n-1)t_{0,1}^2+2(s-1)^2(n+1)(ns-n-s)t_{1,\infty}t_{0,1}}{+s(n+1)(ns-n-s)^2t_{1,\infty}^2}\Bigg)} {2s((n-1)s-n)^2t_{1,\infty}^2t_{0,\infty}^2}
\end{align}
\begin{align}
&S_{1,1}=\frac{2\big(3(s-1)(ns-n-s-1)t_{1,\infty}+((s-1)^2(n^2-1)-3(ns-n-1))\big)}{3t_{1,\infty}t_{0,1}(s-1)(ns-n-s-1)}  \\
&S_{2,1}=\frac{1}{4(ns-n-s-1)(s-1)^2t_{1,\infty}^2t_{0,1}^2}  \\
& \qquad \times \Bigg(12(s-1)^2(ns-n-s-1)t_{1,i}^2+8(s-1)((s-1)^2(n^2-1)-3(ns-n-1))t_{\infty,0}t_{1,\infty} \nn  \\
& \qquad \qquad +\left((s-1)^3(n^2-1)(n+2)-4(s-1)^2(n+1)(n+2)+8s(ns-n-s-2)\right)t_{\infty,0}^2\Bigg) \nn \\
& \qquad \qquad + \frac{(z_0-z_1)}{2(z_0-z_\infty)}\frac{(t_0-t_\infty)^2}{(t_1-t_\infty)^2(t_0-t_1)^2}\frac{n(ns-n-s-1)}{(s-1)} \nn
\end{align}
\begin{align}
&S_{1,s}=\frac{2}{3}\frac{\Bigg(\splitfrac{s(ns-n-s)\big((s-1)^2(n^2-1)-3(ns-n-1)\big)t_{1,\infty}t_{0,\infty}}{-3(s-1)(ns-n-1)(ns-n-s-1)t_{0,1}t_{s,\infty}}\Bigg)}{(ns-n-1)(ns-n-s-1)(s-1)t_{s,\infty}^2t_{0,1}}  \\
&S_{2,s}= \frac{1}{4(ns-n-1)^2(ns-n-s-1)(s-1)^2t_{s,\infty}^2t_{1,0}^2}\times   \\
&  \Bigg( s^2\Big(-8+16n(s-1)-4n(3n-1)(s-1)^2 \nn \\
&\qquad \qquad \qquad \qquad+(n-1)(3n^2+n+2)(s-1)^3\Big)(ns-n-s)^2t_{1,\infty}^2t_{0,\infty}^2  \nn \\
&\qquad -8s(s-1)(ns-n-1)(ns-n-s)((s-1)^2(n^2-1)-3ns+3n+3)t_{0,\infty}t_{s,i\infty}t_{0,1}t_{1,\infty} \nn \\
&\qquad 12(s-1)^2(ns-n-s-1)(ns-n-1)^2t_{s,\infty}^2t_{0,1}^2 \Bigg) \nn
\end{align}

%\vspace*{5cm}

\end{document}